\documentclass[nature,twocolumn,floatfix,superscriptaddress,aps,longbibliography]{revtex4-1} 
\pdfoutput=1
\usepackage{amsfonts,amsmath,amssymb,color,times,graphicx}
\definecolor{darkblue}{rgb}{0, 0, 0.8}
\usepackage[colorlinks=true, breaklinks=true, linkcolor=red, citecolor=darkblue, urlcolor=darkblue]{hyperref}
\usepackage{graphicx}
\usepackage{bmpsize}
\usepackage{amsfonts}
\usepackage{gensymb}
\usepackage{braket}
\usepackage{mathtools}
\usepackage{bm}

\setcitestyle{super}
\graphicspath{{Figs//}}

\begin{document}
\title{Complete field-induced spectral response of the spin-1/2 triangular-lattice antiferromagnet CsYbSe$_2$}

\author{Tao Xie{\color{blue}{$^{\S}$}}}
\thanks{Corresponding author: xiet69@mail.sysu.edu.cn}
\thanks{\\{\color{blue}{$^{\S}$}}~These authors contributed equally to this work}
\affiliation{Neutron Scattering Division, Oak Ridge National Laboratory, Oak Ridge, TN 37831, USA}
\author{A.~A.~Eberharter{\color{blue}{$^{\S}$}}}
\affiliation{Institut f\"{u}r Theoretische Physik, Universit\"{a}t Innsbruck, Innsbruck, Austria}
\author{Jie Xing}
\affiliation{Materials Science and Technology Division, Oak Ridge National Laboratory, Oak Ridge, Tennessee 37831, USA}
\author{S.~Nishimoto}
\affiliation{Department of Physics, Technical University Dresden, 01069 Dresden, Germany}
\affiliation{Institute for Theoretical Solid State Physics, IFW Dresden, 01069 Dresden, Germany}
\author{M.~Brando}
\affiliation{Max Planck Institute for Chemical Physics of Solids, N\"{o}thnitzer Str. 40, D-01187 Dresden, Germany}
\author{P.~Khanenko}
\affiliation{Max Planck Institute for Chemical Physics of Solids, N\"{o}thnitzer Str. 40, D-01187 Dresden, Germany}
\author{J.~Sichelschmidt}
\affiliation{Max Planck Institute for Chemical Physics of Solids, N\"{o}thnitzer Str. 40, D-01187 Dresden, Germany}
\author{A. A. Turrini}
\affiliation{Laboratory for Neutron Scattering and Imaging, Paul Scherrer Institut, CH-5232 Villigen-PSI, Switzerland}
\author{D. G. Mazzone}
\affiliation{Laboratory for Neutron Scattering and Imaging, Paul Scherrer Institut, CH-5232 Villigen-PSI, Switzerland}
\author{P. G. Naumov}
\affiliation{Quantum Criticality and Dynamics Group, Paul Scherrer Institut, CH-5232 Villigen-PSI, Switzerland}
\affiliation{Orange Quantum Systems B.V., Elektronicaweg 22628 XG DelftThe Netherlands}
\author{L.~D.~Sanjeewa}
\affiliation{Materials Science and Technology Division, Oak Ridge National Laboratory, Oak Ridge, Tennessee 37831, USA}
\author{N.~Harrison}
\affiliation{National High Magnetic Field Laboratory, Los Alamos National Laboratory, Los Alamos, New Mexico 87545, USA}
\author{Athena~S.~Sefat}
\affiliation{Materials Science and Technology Division, Oak Ridge National Laboratory, Oak Ridge, Tennessee 37831, USA}
\author{B.~Normand}
\affiliation{Laboratory for Theoretical and Computational Physics, Paul Scherrer Institut, CH-5232 Villigen-PSI, Switzerland}
\affiliation{Institute of Physics, Ecole Polytechnique F\'ed\'erale de Lausanne (EPFL), CH-1015 Lausanne, Switzerland}
\author{A. M. L\"auchli}
\thanks{Corresponding author: andreas.laeuchli@psi.ch}
\affiliation{Laboratory for Theoretical and Computational Physics, Paul Scherrer Institut, CH-5232 Villigen-PSI, Switzerland}
\affiliation{Institute of Physics, Ecole Polytechnique F\'ed\'erale de Lausanne (EPFL), CH-1015 Lausanne, Switzerland}
\author{A.~Podlesnyak}
\affiliation{Neutron Scattering Division, Oak Ridge National Laboratory, Oak Ridge, TN 37831, USA}
\author{S.~E.~Nikitin}
\thanks{Corresponding author: stanislav.nikitin@psi.ch}
\affiliation{Quantum Criticality and Dynamics Group, Paul Scherrer Institut, CH-5232 Villigen-PSI, Switzerland}

\begin{abstract}
Fifty years after Anderson's resonating valence-bond proposal, the spin-1/2 triangular-lattice Heisenberg antiferromagnet (TLHAF) remains the ultimate platform to explore highly entangled quantum spin states in proximity to magnetic order. Yb-based delafossites are ideal candidate TLHAF materials, which allow experimental access to the full range of applied in-plane magnetic fields. We perform a systematic neutron scattering study of CsYbSe$_2$, first proving the Heisenberg character of the interactions and quantifying the second-neighbour coupling. We then measure the complex evolution of the excitation spectrum, finding extensive continuum features near the 120$\degree$-ordered state, throughout the 1/3-magnetization plateau and beyond this up to saturation. We perform cylinder matrix-product-state (MPS) calculations to obtain an unbiased numerical benchmark for the TLHAF and spectacular agreement with the experimental spectra. The measured and calculated longitudinal spectral functions reflect the role of multi-magnon bound and scattering states. These results provide valuable insight into unconventional field-induced spin excitations in frustrated quantum materials.
\end{abstract}

\maketitle

\bigskip
\noindent
{\bf INTRODUCTION}
\smallskip

\noindent
Frustrated quantum magnets provide an intriguing playground for investigating novel many-body phenomena in condensed matter~\cite{Savary2017,broholm2020quantum}. The triangular-lattice Heisenberg antiferromagnet (TLHAF) is a prototypical example of geometrical frustration, and its ground state in the quantum limit of $S = 1/2$ spins has the 120$\degree$ AF order of the classical (large-$S$) case \cite{Starykh2015}, albeit with an ordered moment strongly suppressed by quantum fluctuation effects (\cite{li2022magnetization} and references therein). Proposals to capture these effects include the resonating valence-bond (RVB) paradigm~\cite{Anderson}, and the addition of a weak next-neighbour HAF interaction ($0.06 \lesssim J_2/J_1 \lesssim 0.15$) does drive the $S = 1/2$ TLHAF into a quantum spin-liquid (QSL) phase \cite{Kaneko2014,Li2015,Zhu2015,Hu2015} of some type \cite{Hu2019,Sherman2022,Drescher2022}. Over a finite range of applied magnetic fields, AF quantum fluctuations favour a collinear up-up-down (UUD) ordered phase and thus stabilize a magnetization plateau with $M = M_{\mathrm{Sat}}/3$ (where $M_{\mathrm{Sat}}$ is the saturation magnetization) \cite{chubokov1991quantum, yamamoto2014quantum}.

Theoretical research on the TLHAF has been driven by new generations of TL materials. The Cs$_2$Cu$X_4$ compounds ($X$ = Cl, Br) \cite{Coldea2001,Tanaka2002} inspired studies of spatially anisotropic TLs \cite{Starykh2015}. The low-spin cobaltates Ba$_3$Co$X_2$O$_9$ ($X$ = Nb, Sb) \cite{Shirata2012,lee2014series,Ma2016,Ito2017,Macdougal2021} and Ba$_8$CoNb$_6$O$_{24}$ \cite{Cui2018} focused attention on XXZ spin anisotropy \cite{yamamoto2014quantum}. The first 4$f$ TLAF, YbMgGaO$_4$ \cite{li2015rare,paddison2017continuous}, sparked more extensive studies of spin anisotropy that enriched the phase diagram and revealed the connection to the QSL phase of the $J_1$-$J_2$ TLHAF \cite{Liyaodong2016,zhu2018topography,maksimov2019anisotropic}.

\begin{figure*}[t]
\center{\includegraphics[width=1\linewidth]{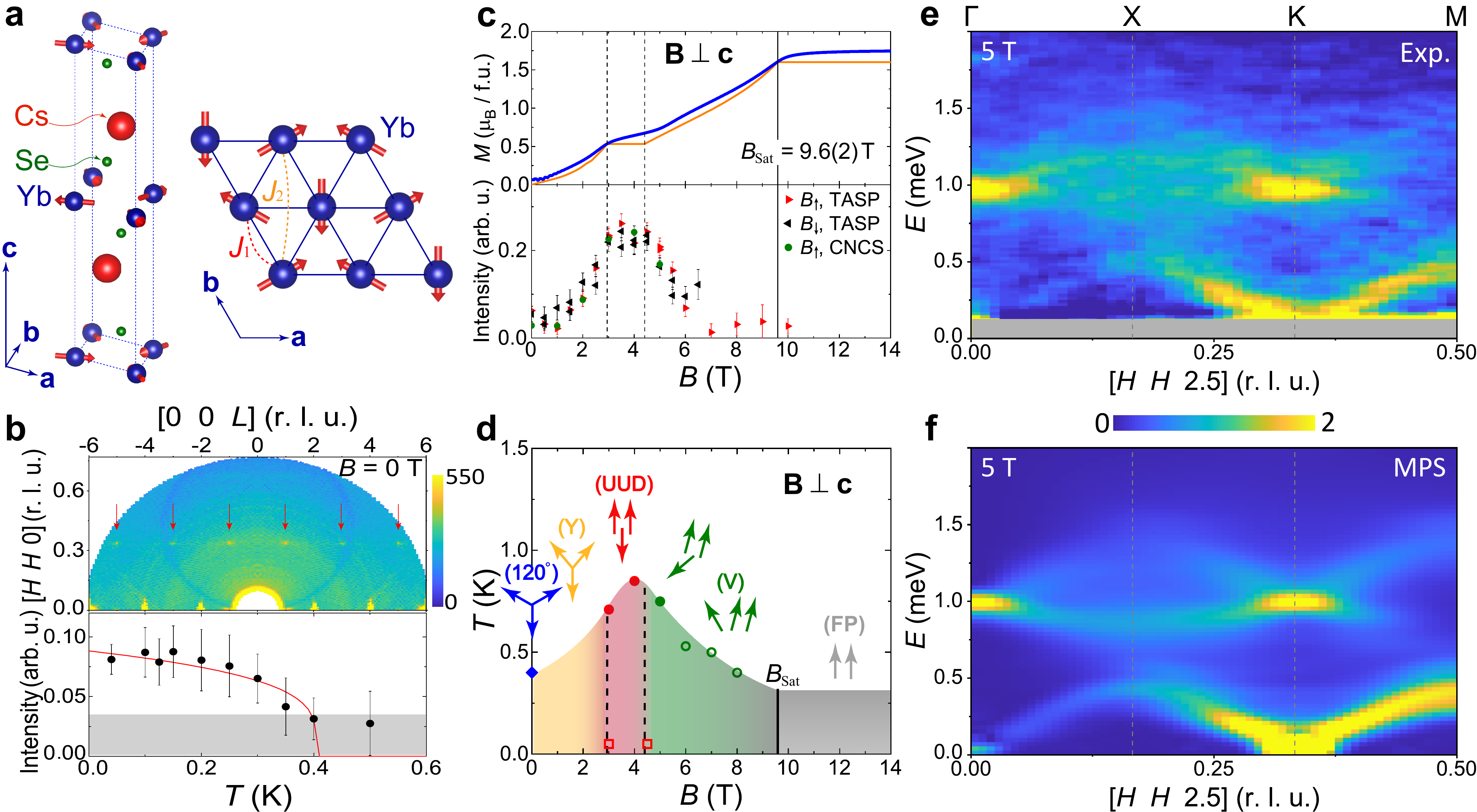}}
\caption{{\bf Structure, properties, phase diagram and finite-field spectra of CsYbSe$_2$.} \textbf{a} Crystal structure of CsYbSe$_2$ and representation of the ideal Yb$^{3+}$ TL layer. The red arrows represent the ordered spins of the weak 120$\degree$ AF order at zero field.
\textbf{b} Upper panel: elastic scattering intensity in the $(H H L)$ plane for 0~T, with the data presented as described in Supplementary Note 2B. Red arrows indicate weak magnetic intensity peaks at $\mathbf{Q} = (1/3, 1/3, \pm{}L)$ with $L = 1,3,5$. Lower panel: temperature-dependence of the $(1/3, 1/3, 1)$ peak area at zero field. The red solid line is a fit to an order-parameter form. The grey shaded area represents the approximate sensitivity limit of our measurement.
\textbf{c} Upper panel: isothermal magnetization (blue symbols) measured at $T = 0.4$~K as a function of magnetic field applied in the $ab$ plane and with the van Vleck contribution subtracted as described in Supplementary Note 1C. The solid orange line shows a grand canonical density-matrix renormalization-group (DMRG) calculation of the magnetization performed for the TLHAF using parameters deduced in Fig.~\ref{camea}, from which we determined the saturation field, $B_{\mathrm{Sat}} = 9.6(2)$ T (vertical solid line), and the lower and upper boundaries of the 1/3 plateau as $B_{\rm l} = 2.95(6)$ T and $B_{\rm u} = 4.5(1)$ T (vertical dashed lines). Lower panel: integrated intensity of the $(1/3, 1/3, 1)$ magnetic Bragg peak measured at $T < 0.05$~K.
\textbf{d} Phase diagram of CsYbSe$_2$. The blue diamond is the phase-transition temperature obtained from neutron diffraction at zero field. The two open squares indicate the region (3~T~$\le B \le$~4.5~T) where the peak intensity in the lower half of panel \textbf{c} remains almost unchanged. The solid and open circles represent respectively the temperatures of sharp peaks and broad humps in the corresponding specific-heat curves, as described in Supplementary Note 1D. The arrows indicate schematically the spin order of the five phases of the classical TLHAF, which are consistent with the magnetic peaks we observe.
\textbf{e} INS spectrum measured under a magnetic field of 5~T, showing an absence of well defined $\Delta S = 1$ excitations away from the $\Gamma$ and K points but extensive continuum features (Fig.~\ref{Dispersion}). The grey horizontal bar at low energy masks the elastic-line contribution.
\textbf{f} Comparison with a matrix-product-state (MPS) calculation of the spectrum at the same field.}
\label{fig1}
\end{figure*}

In this context, the Yb-based delafossite family $A$Yb$Q_2$, with $A$ an alkali metal and $Q$ a chalcogenide, has attracted widespread attention \cite{liu2018rare,Schmidt2021}. The Yb ions form perfect and well separated TLs (Fig.~\ref{fig1}a) \cite{baenitz2018naybs,ding2019gapless,xing2021synthesis} without the structural disorder intrinsic to YbMgGaO$_4$ \cite{zhu2017disorder,li2017crystalline}. The combination of strong spin-orbit coupling and the crystalline electric field (CEF) creates a ground-state doublet that gives an effective $S = 1/2$ pseudospin at low temperatures \cite{Rau2018Frustration,baenitz2018naybs,ding2019gapless,Zhang2021CEF}. Although the $J = 7/2$ CEF level structure is manifest in a strong spatial anisotropy of the response to applied magnetic fields \cite{Ranjith2019naybse,Pocs2021}, initial scattering studies provided no evidence for a strongly non-Heisenberg pseudospin Hamiltonian \cite{bordelon2019field,Xing20192}. Early specific heat, magnetization, muon spin-rotation spectroscopy and neutron diffraction studies of multiple $A$Yb$Q_2$ materials found no magnetic order at zero field down to their base temperatures \cite{Ranjith2019naybse,bordelon2019field,Xing20192,ding2019gapless,Sarkar2019}, but recent studies, including our own (Fig.~\ref{fig1}b), indicate its presence in some materials at the 0.1 K scale. The 1/3-magnetization plateau is found at in-plane fields in the 3-5 T range \cite{Ranjith2019naybse,bordelon2019field,Xing20192,ranjith2019field} (Fig.~\ref{fig1}c), with robust UUD order up to 1 K. Clearly the delafossite family offers an excellent platform to study the field-controlled magnetic states of the $S = 1/2$ TLAF.

Initial inelastic neutron scattering (INS) measurements on single-crystalline Yb delafossites at zero field \cite{Xing20192,dai2020spinon} suggested a gapless excitation continuum, which was interpreted as originating from a QSL ground state, but appears to persist even in the presence of weak magnetic order \cite{Scheie2021}. Early INS studies of the spin dynamics in the field-induced phases were limited by their polycrystalline samples~\cite{bordelon2019field,bordelon2020spin}, but the 1/3-magnetization plateau has recently been analyzed in some detail \cite{Scheie2022}. As in Ba$_3$CoSb$_2$O$_9$, where the plateau has been reached despite the higher energy scales in this family of materials \cite{kamiya2018nature}, the magnetic excitations were captured largely by semiclassical nonlinear spin-wave theory (SWT). The lower in-plane energy scale and vanishing inter-plane coupling in delafossites present more experimental challenges, but also the key advantage of reaching saturation within laboratory-available magnetic fields.

On the theoretical side, the challenge of computing the dynamical spectral functions of frustrated models lies in the absence of analytical methods that capture all the physics of non-collinear magnetic states with a field-controlled ratio of weak order to strong quantum fluctuations. The application of unbiased numerical methods, meaning those whose truncation methodology can be extended systematically to convergence, has in the past been impossible, but continuous progress in dynamical quantum Monte Carlo techniques and matrix-product-state (MPS) representations is placing this goal within reach. For the TLHAF, the zero-field spectral function has been obtained by a number of biased methods, by which we mean those based on initial assumptions that have to be assesed {\it a posteriori}; these include series expansions~\cite{Zheng2006}, interacting spin waves~\cite{Chernyshev2009,Mourigal2013}, Schwinger bosons \cite{Ghioldi2018,Zhang2022}, bond operators \cite{Syromyatnikov2022a} and variational Monte Carlo~\cite{Ferrari2019}. Despite recent progress with MPS calculations in a cylinder geometry~\cite{Sherman2022,Drescher2022}, the full field-induced dynamics has remained an unsolved problem. All of these methods produce scattering continua whose origin may lie in fractional excitations, multi-magnon states or possibly neither.

In this work we perform high-resolution neutron spectroscopy on CsYbSe$_2$ using two different spectrometers to span the full range of applied in-plane fields, meaning from zero to beyond saturation. Our measurements reveal the pronounced changes in the magnetic excitation spectrum as it evolves with the magnetic field, and we associate these with the field-driven phase transitions of the ground state (Fig.~\ref{fig1}d). In parallel we perform large-cylinder MPS calculations of the full TLHAF spectral function at all fields to obtain a hitherto unavailable benchmark for the model, semi-quantitative agreement with experiment (Figs.~\ref{fig1}e-f) and a robust foundation for any effective quasiparticle descriptions of the spin dynamics.

\begin{figure*}[t]
\center{\includegraphics[width=0.65\linewidth]{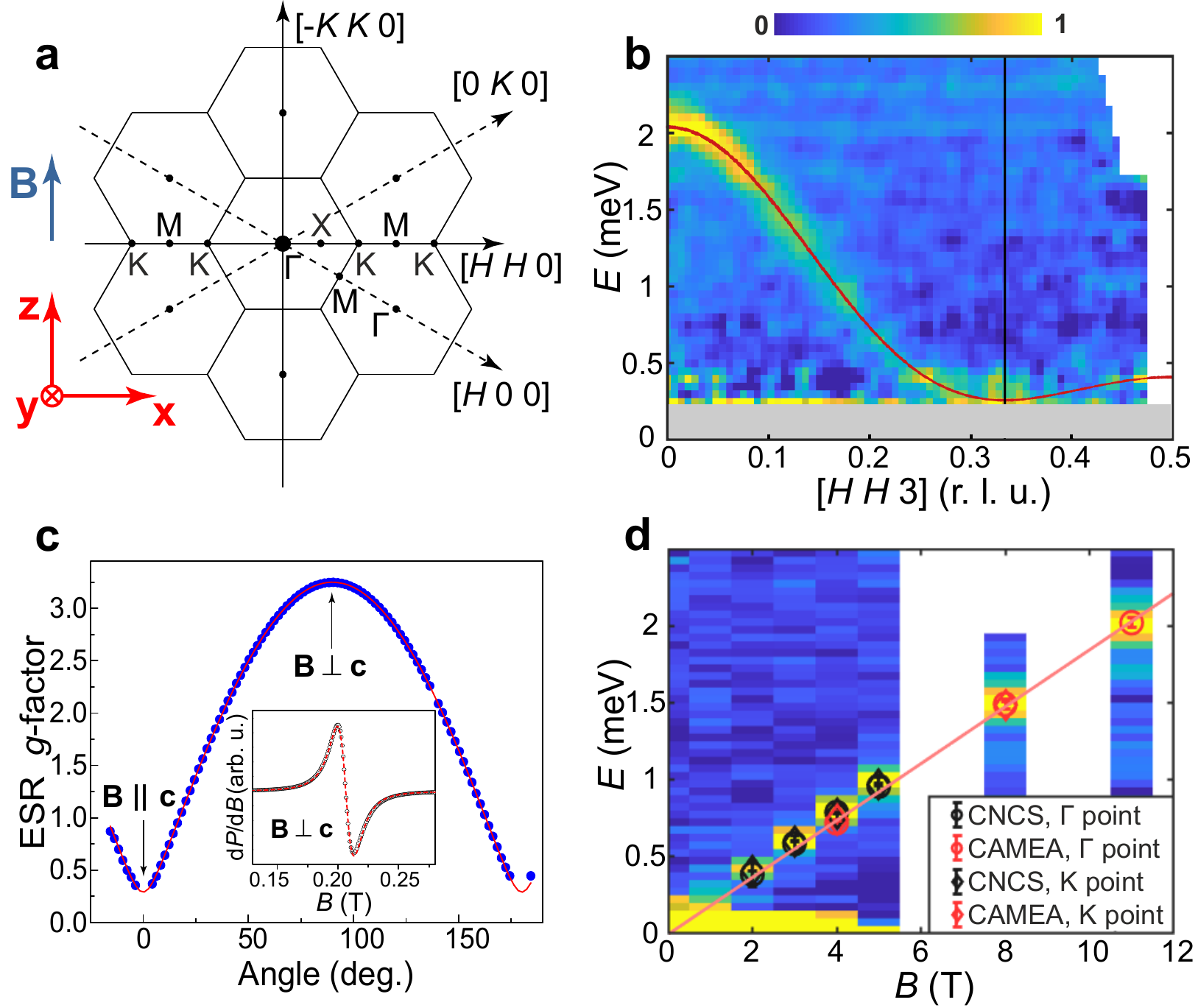}}
\caption{{\bf INS and ESR determination of the spin Hamiltonian.}
\textbf{a} Representation of the Brillouin zone in the $(H~K~0)$ plane and definition of directions ${\hat x} \parallel\ [1~1~0]$, ${\hat y} \parallel\ [0~0~1]$ and ${\hat z} \parallel \mathbf{B}$, which is orthogonal to ${\hat x}$ and ${\hat y}$.
\textbf{b} Spin excitations along the $[H~H~3]$ direction measured on CAMEA at $T = 0.05$~K in the field-polarized regime ($B = 11$~T). The solid line shows the dispersion calculated using linear spin-wave theory (SWT) with $J_1 = 0.395(8)$~meV, $J_2 = 0.011(4)$~meV and $g = 3.2$. The grey horizontal bar at low energy masks the elastic-line contribution.
\textbf{c} Angular dependence of the $g$-factor measured by ESR at $T = 15$~K. The solid line indicates the form $g(\theta) = \sqrt{g_{ab}^2 \sin^2{\theta} + g_{c}^2 \cos^2{\theta}}$, which allows the extraction of the strongly anisotropic in- and out-of-plane coefficients $g_{ab} = 3.25$ and $g_c = 0.3$. Inset: representative ESR spectrum with a Lorentzian fit shown by the dashed red line.
\textbf{d} Field-dependence of the INS signal at the $\Gamma$ point; data for $B \leq\ 5$~T were measured on CNCS and data for $B = 8$ and 11~T on CAMEA. Black and red points show the respective positions of the magnon mode as extracted from the CNCS and CAMEA datasets, for both the $\Gamma$ and K points. The solid line shows a linear fit that yields $g_{ab} = 3.20(6)$.}
\label{camea}
\end{figure*}

\bigskip
\noindent
{\bf RESULTS}
\smallskip

\noindent
{\bf {Ground state at zero field}}
\smallskip

\noindent
The growth and structural characterization of our single crystals are summarized in the Methods section and detailed in Supplementary Note 1A. Neutron diffraction at zero magnetic field (Supplementary Note 2) reveals a series of weak magnetic intensity peaks at $\mathbf{Q} = (1/3, 1/3, L)$ for odd-integer $L$ (Fig.~\ref{fig1}b). The $(1/3, 1/3, 1)$ peak develops a finite intensity below $T \simeq 0.4$~K, which increases on cooling. The propagation wavevector matches the 120$\degree$ state of the TLHAF with AF out-of-plane correlations (represented by the arrows in Fig.~\ref{fig1}a) and the low-temperature ordered moment is $m_{\mathrm{Yb}} \simeq 0.1~\mu_{\mathrm{B}}$. In Supplementary Note 2B we extract the in- and out-of-plane correlation lengths, $\xi_{ab} = 60(7)$~{\AA} and $\xi_c = 23(5)$~{\AA}, which are not resolution-limited, meaning that CsYbSe$_2$ does not exhibit true, long-ranged AF order at zero field down to $T = 0.02$~K. However, the presence of the magnetic peak clearly excludes a QSL, as in KYbSe$_2$~\cite{Scheie2021} but in contrast to NaYbSe$_2$ \cite{dai2020spinon}. Given that CsYbSe$_2$ (space group P$6_3/mmc$) has AA layer stacking (Fig.~\ref{fig1}a), which should favour an unfrustrated collinear $c$-axis order, whereas the Na, K and Rb materials (space group R$\overline{3}m$) have an ABC stacking that should produce interlayer frustration, we suggest in Supplementary Note 1B that the origin of this behaviour may instead lie in the next-nearest neighbour coupling, $J_2$ (below).

\medskip
\noindent
{\bf {Magnetic phase diagram of CsYbSe$_2$}}
\smallskip

\noindent
We performed low-temperature magnetization, specific-heat and neutron diffraction measurements over a wide field range, as described in Supplementary Notes 1 and 2. Figure~\ref{fig1}c shows isothermal magnetization data, with evidence of a plateau at $M_{\mathrm{Sat}}/3$ corresponding to the UUD phase~\cite{chubokov1991quantum,yamamoto2014quantum}. To interpret these data despite their finite-temperature rounding, we estimate $B_{\mathrm{Sat}} = 9.6(2)$~T from the TLHAF model parameters obtained by high-field INS (Fig.~\ref{camea}) and perform a grand canonical DMRG calculation~\cite{Hotta2013} of the magnetization (Supplementary Note 4) that allows us to deduce the boundaries of the 1/3-magnetization plateau. We compare these data with the integrated intensity of the $(1/3, 1/3, 1)$ magnetic peak, which increases strongly from 0~T to the UUD state, then remains maximal and almost constant over the field range of the plateau (3~T~$\le B \le 4.5$~T), before decreasing strongly towards the fully polarized (FP) state. Our thermodynamic and neutron diffraction results yield the field-temperature phase diagram shown in Fig.~\ref{fig1}d, where we indicate the spin alignments of the classical TLHAF.

\begin{figure*}[t]
\center{\includegraphics[width=1\linewidth]{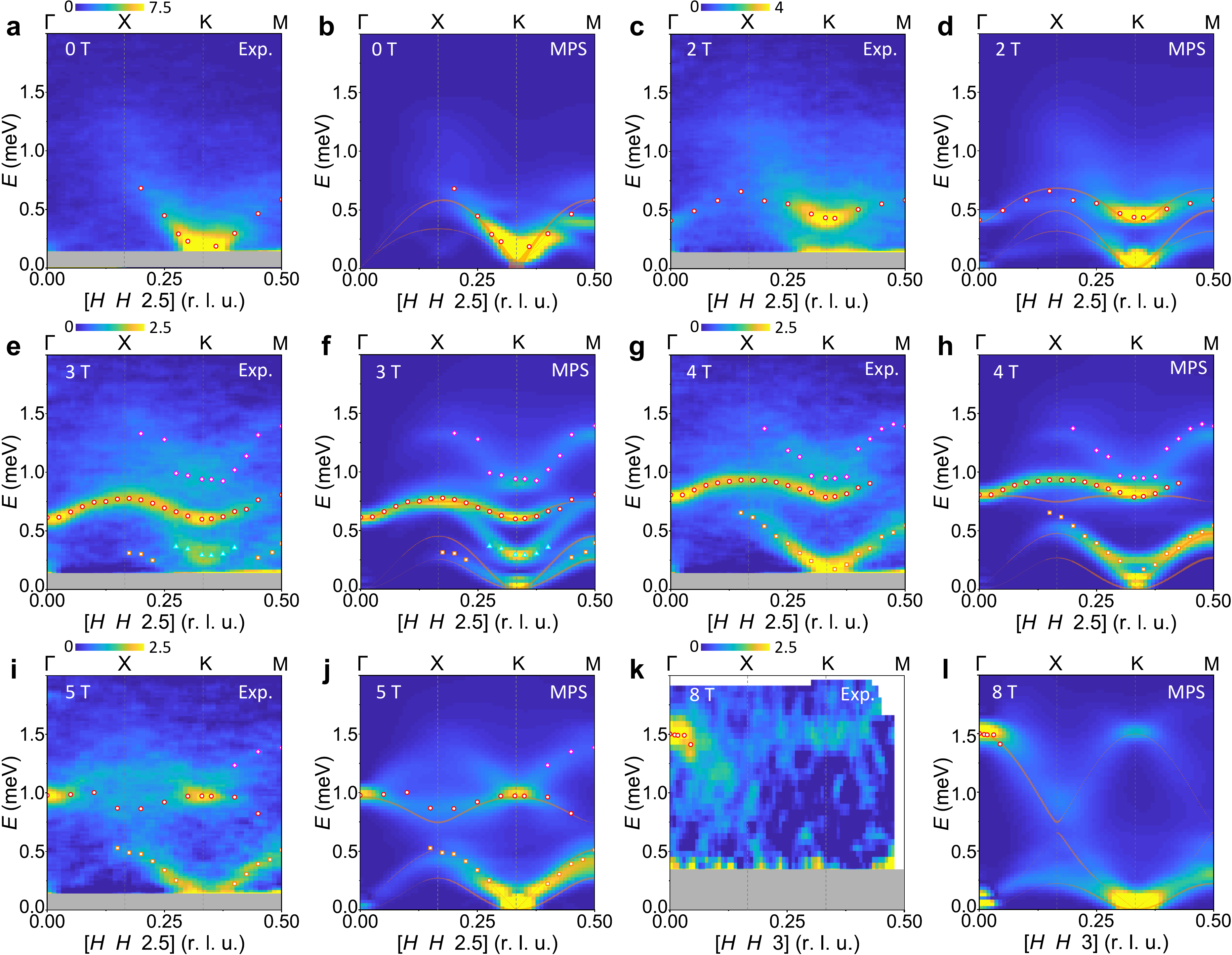}}
\caption{{\bf Complete field-induced spectral response of CsYbSe$_2$.}
Panels \textbf{a}, \textbf{c}, \textbf{e}, \textbf{g}, \textbf{i} and \textbf{k} show spin excitation spectra measured under different magnetic fields at $T$ = 0.07~K. The open circles, squares, triangles and diamonds indicate respectively excitation features in the categories I, II, III and IV described in the text. All data have been symmetrized according to the crystal symmetry. The orthogonal in-plane integration range along the $[-K~K~0]$ direction is $K = [-0.05, 0.05]$ and the out-of-plane range is $1.2 \le L \le 3.8$ for our CNCS data (0-5 T) and $2 \le L \le 4$ on CAMEA (8 T). The background subtraction is described in Supplementary Note 2E. The narrow horizontal grey regions mask the elastic line. Panels \textbf{b}, \textbf{d}, \textbf{f}, \textbf{h}, \textbf{j} and \textbf{l} present dynamical spin structure factors obtained at different magnetic fields by cylinder MPS calculations (Supplementary Note 5). Colour bars represent both the measured and calculated intensities in a single set of arbitrary units (i.e.~the same units are used at all fields). Orange lines show the mode positions and intensities given by linear SWT with the same interaction parameters. The open points are identical to those shown in panels \textbf{a}, \textbf{c}, \textbf{e}, \textbf{g}, \textbf{i} and \textbf{k}. }
\label{Dispersion}
\end{figure*}

\medskip
\noindent
{\bf {Parameters of the magnetic Hamiltonian}}
\smallskip

\noindent
We made INS measurements up to 5~T on the time-of-flight (ToF) spectrometer CNCS at ORNL and up to 11~T on the multiplexing spectrometer CAMEA at PSI, as detailed in the Methods section. To quantify the parameters of the spin Hamiltonian, we exploit our ability to perform INS at fields $B > B_{\mathrm{Sat}}$, where the magnetic excitations of the FP phase can be described by linear SWT. Figure~\ref{camea}b shows that the spectrum measured along $[H~H~3]$ at 11~T consists of a single, sharp magnon mode with a cosinusoidal dispersion above a field-induced gap at the K point. In Supplementary Note 2D we show CNCS data indicating a complete lack of out-of-plane dispersion over the whole field range, and hence that a two-dimensional TL model is appropriate. We fit this dispersion using the \textsc{SpinW} package \cite{spinw} by considering a Hamiltonian with an anisotropic XXZ-type $J_1$ term and a Heisenberg $J_2$ term (Supplementary Note 3A). With the in-plane $g$-factor fixed (below), the optimal fit (solid red line in Fig.~\ref{camea}b) yields two essential pieces of information. First, the nearest-neighbour interaction has no XXZ anisotropy within the precision of the measurement, i.e.~despite the strongly anisotropic field response \cite{Schmidt2021,Pocs2021}, the spin dynamics are of Heisenberg type; in Ref.~\cite{Rau2018Frustration} it was shown how these contrasting forms of behaviour can appear simultaneously in edge-sharing octahedral Yb$^{3+}$ systems. Second, the next-neighbour interaction is sufficiently weak, $J_2/J_1 \simeq 0.03$, that CsYbSe$_2$ remains on the ordered side of the phase boundary separating the 120$^{\degree}$ and QSL states in the $S = 1/2$ $J_1$-$J_2$ TLHAF~\cite{Kaneko2014,Li2015,Zhu2015,Hu2015}. We therefore conclude that a $J_1$-$J_2$ Heisenberg Hamiltonian
\begin{eqnarray}
\hspace{-12pt} \mathcal{H} = J_1 \sum_{\langle i,j \rangle} \mathbf{S}_i \cdot \mathbf{S}_j + J_2 \sum_{\langle\langle i,j  \rangle\rangle} \mathbf{S}_i \cdot \mathbf{S}_j - \mu_B g_{ab} B \sum_i{S}^z_i,
\label{Eq:Ham}
\end{eqnarray}
with $J_1 = 0.395$ meV, provides a complete description of the low-energy magnetic behaviour in CsYbSe$_2$.

Turning to fields below saturation, we begin in Fig.~\ref{camea}d by considering constant-$\mathbf{Q}$ cuts at the $\Gamma$-point for each field. These show a clear, single-peak feature for $B \geq 2$~T, whose energy obeys the field-linear form $\hbar\omega(B) = \hbar\omega_0 + g_{ab} \mu_{\mathrm{B}}BS$ with $g_{ab} = 3.20(6)$ and $\hbar\omega_0 = 0.00(2)$~meV. This field-induced behaviour at $\Gamma$ is generic for the Heisenberg model~\cite{Oshikawa2002}, reinforcing our conclusion concerning the absence of XXZ anisotropy, and in the TLHAF is also present at K (Fig.~\ref{camea}d). To characterize the anisotropic field response, we have performed electron spin resonance (ESR) measurements that determine the $g$-tensor parameters shown in Fig.~\ref{camea}c. The narrow and well-defined ESR spectrum reflects the high quality of our crystal. The best fit in Fig.~\ref{camea}c, $g_{ab} = 3.25(0)$ and $g_c = 0.3(0)$, completes our determination of the Hamiltonian parameters for CsYbSe$_2$ in any applied field and also shows the consistency of our INS result for $g_{ab}$. Our measurements also demonstrate that $g_{ab}$ is isotropic in the $ab$ plane to within the experimental accuracy (data not shown). The very small $g_c$ is a consequence of strong hybridization with the first excited CEF doublet \cite{Pocs2021}, and we comment below on its role in our scattering study.

\medskip
\noindent
{\bf {MPS calculations of spectral functions}}
\smallskip

\noindent
To interpret the measured spin dynamics, we have performed MPS calculations on a finite cylinder to obtain the dynamical spectral function of the TLHAF with $J_2 = 0.03 J_1$, where the energy unit is fixed to $J_1 = 0.395$ meV. The cylinder size, matrix bond dimensions and time-evolution procedures are summarized in the Methods section and their convergence to the properties of the TLHAF is benchmarked in Supplementary Note 5. We compute the spin correlation functions both transverse and longitudinal to the applied field, which in the notation of Fig.~\ref{camea}a are respectively $S_{xx} ({\mathbf Q}, \omega)$ and $S_{zz} ({\mathbf Q}, \omega)$. In experiment, the strong $g$-tensor anisotropy ($g_{ab} \gg\ g_c$) means that the component fluctuating parallel to the $c$ axis ($S_{yy}$) is hidden [Fig.~\ref{camea}a defines the $(HKL)$ and $(xyz)$ coordinate frames]. The measured intensities then represent the sum of $S_{xx}$ and $S_{zz}$ weighted by the polarization factor, which ensures that spectra taken along $[H~H~0]$ have no contribution from $S_{xx}$, i.e.~only from the component longitudinal to the applied field. In order to sample both components, in Fig.~\ref{Dispersion} we integrate our INS and MPS spectra over a wide range of the out-of-plane momentum, $L$.

\medskip
\noindent
{\bf {Field-induced evolution of the spectrum}}
\smallskip

\noindent
As expected from the phase diagram (Fig.~\ref{fig1}d), both the observed and computed spectra in Fig.~\ref{Dispersion} are readily classified by their field-induced evolution into four regimes, to which we refer as Y, UUD, V and FP (the last analysed in Fig.~\ref{camea}). Starting with Y, we have shown (Fig.~\ref{fig1}b) that the zero-field ground state of CsYbSe$_2$ is consistent with three-sublattice 120$^{\circ}$ order, and thus the spectrum should contain three excitation branches. However, both the INS and MPS spectra exhibit only a broad, V-shaped continuum around K (Figs.~\ref{Dispersion}a-b), similar to the spectra observed in NaYbSe$_2$~\cite{dai2020spinon} and KYbSe$_2$~\cite{Scheie2021}. Because this clear signature of strong quantum fluctuations on top of weak magnetic order has received extensive theoretical \cite{Chernyshev2009,Ghioldi2018,Ghioldi2022} and numerical analysis \cite{Zheng2006,Ferrari2019,Scheie2021,Sherman2022,Drescher2022}, which our results confirm but do not extend, we focus rather on adding to the understanding of the finite-field spectra.

At 2~T (Figs.~\ref{Dispersion}c-d), most of the spectral intensity shifts upwards, forming the broad feature, with a gap around 0.4~meV at $\Gamma$ and K, seen in Fig.~\ref{camea}d. Away from $\Gamma$ and K, however, it has no well defined magnonic form and the spectrum appears as a weak and highly dispersed continuum. Nevertheless, we mark the maximum intensity of this feature by the circles in Figs.~\ref{Dispersion}c-d and refer to it as mode I, observing its bandwidth falling rapidly across the Y regime. Another mode is present at lower energies, whose gapless nature is clearly visible in the MPS spectrum. Linear SWT captures rather well the maximum of mode I, and also finds two gapless branches, but cannot reproduce the extreme broadening of the measured modes away from $\Gamma$ and K, their intensity distribution or the continuum scattering above mode I.

\begin{figure}[t]
\center{\includegraphics[width=1\linewidth]{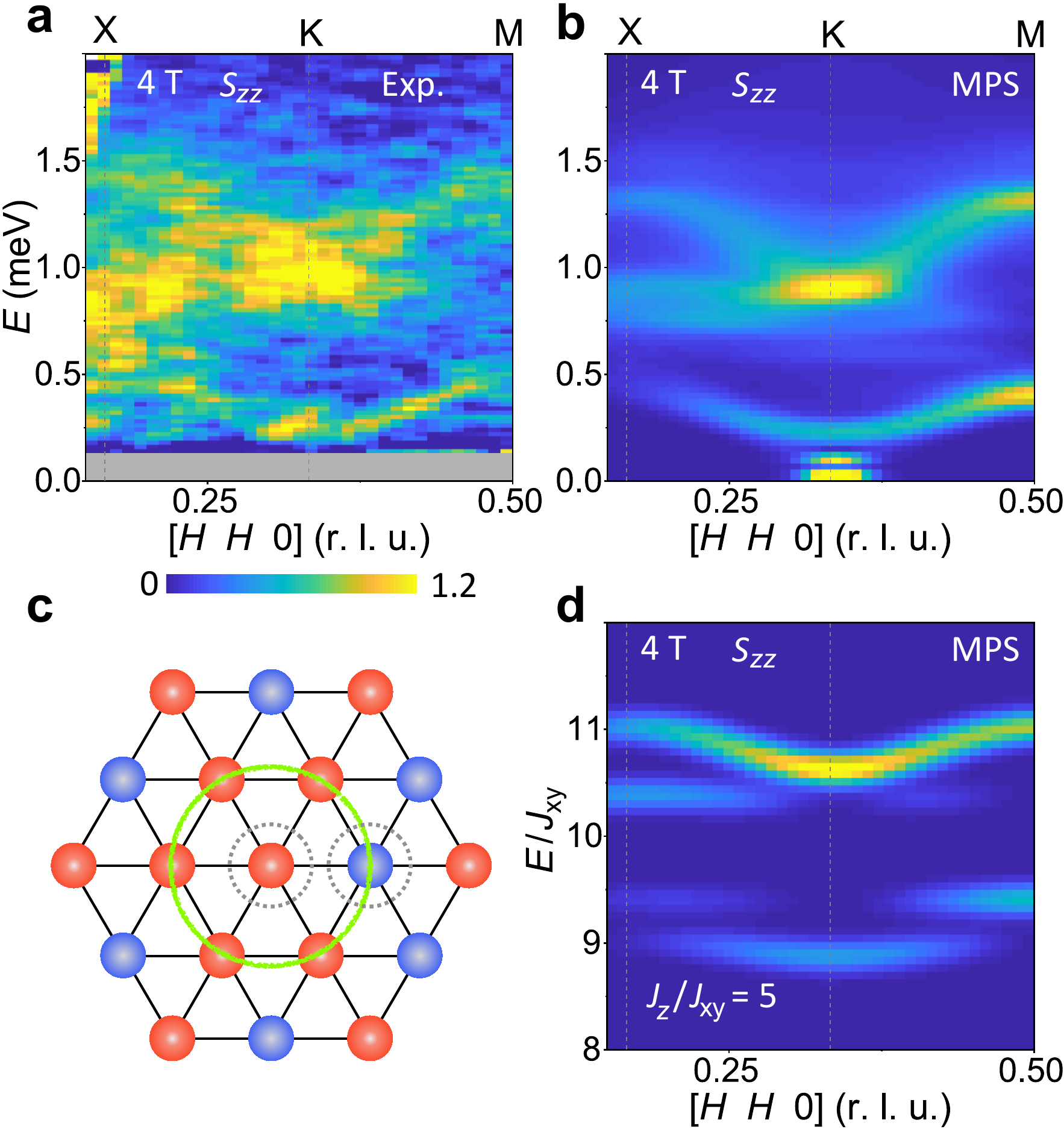}}
\caption{{\bf Longitudinal spin excitations.}
\textbf{a} Longitudinal component of the excitation spectrum extracted for the $[H~H~0]$ direction at $B = 4$ T. The orthogonal in-plane integration range is $K = [-0.05, 0.05]$ and the out-of-plane range is $L = [-0.5, 0.5]$. The narrow grey region masks the elastic line.
\textbf{b} Corresponding MPS calculation of the longitudinal component, $S_{zz} (\mathbf{Q},\omega)$, of the dynamical structure factor.
\textbf{c} Schematic representation of spin-flip processes in the UUD phase: red and blue circles represent respectively U and D spins, the grey dashes highlight flipped spins (U $\rightarrow$ D and D $\rightarrow$ U) and the green circle delineates the hexagon on which the blue flipped spin may propagate at no energy cost in the Ising limit, ensuring its localization. \textbf{d} MPS calculation of $S_{zz} (\mathbf{Q},\omega)$ for the strongly Ising-type parameter choice $J_{z} = 5 J_{xy}$~(Supplementary Note 7); the mode energies are shown in units of $J_{xy}$ to illustrate the role of this interaction in setting the splitting and dispersion of the bound states centred at $2 J_1 = 2 J_z$.}
\label{Szz}
\end{figure}

The data collected at 3 and 4~T represent respectively the lower edge and upper middle of the UUD regime (Fig.~\ref{fig1}c). In contrast to the Y phase, a number of rather sharp excitations extend across the full Brillouin zone, and at 3~T we identify four distinct features (Figs.~\ref{Dispersion}e-f). Mode I shifts upwards, becomes resolution-limited and has intensity over a large $\mathbf{Q}$ range. A weak and very low-lying feature II is visible only around its maximum near 0.4~meV. A broad feature III is concentrated around the K point and disperses upwards to touch mode I. A continuum feature IV disperses from around 0.8~meV at K to 1.3~meV at X and M. At 4~T (Figs.~\ref{Dispersion}g-h), features II and III have almost merged to become a strong, spin-wave-like branch. Mode I continues its upward shift while continuum IV remains almost unchanged in position and intensity. Modes I-III have been observed in Ba$_3$CoSb$_2$O$_9$ \cite{kamiya2018nature}, and very recently all four features were measured in KYbSe$_2$ \cite{Scheie2022}. Both studies used a nonlinear SWT to obtain a good account of modes I-III, and in KYbSe$_2$ it was suggested that feature IV is a two-magnon continuum. Given that the 1/3 plateau is absent in linear SWT, it is not surprising that the orange lines in Figs.~\ref{Dispersion}f and \ref{Dispersion}h provide at best partial agreement with some observed branches. By contrast our MPS calculations provide a quantitatively excellent description of every feature in the observed UUD spectra, which will allow a deeper analysis in Fig.~\ref{Szz}.

Proceeding into the V phase at 5~T causes a qualitative modification of the spectrum (Figs.~\ref{Dispersion}i-j). As in the Y phase, no clear magnon branches are visible away from $\Gamma$ and K. More specifically, mode II softens, broadens and decreases in intensity, while mode III merges fully with it. Mode I decays into a broad continuum with sharp intensity peaks only at $\Gamma$ and K, and obscures continuum IV. Linear SWT traces only the lower boundary of the mode-I continuum and the dispersion of mode II. The 8~T dataset in Fig.~\ref{Dispersion}k shows a sharp band maximum at $\Gamma$ (mode I) together with a weak replica at K. Here our MPS results (Fig.~\ref{Dispersion}l) clarify how mode I becomes very broad and mode II becomes very soft; linear SWT provides an acceptable guide to the positions, but absolutely not to the emerging mid-zone continuum nature, of these features. In Supplementary Note 6 we present cuts through the data displayed in Figs.~\ref{Dispersion}a-j that confirm the near-quantitative agreement between the INS and MPS spectra at almost all points in ${\mathbf Q}$ and $\omega$.

\medskip
\noindent
{\bf {Two-magnon bound and scattering states}}
\smallskip

\noindent
To analyse the spectra in Fig.~\ref{Dispersion} we begin in the UUD phase, where all the ordered moments are orientated (anti)parallel to the field and thus the $S_{zz}$ channel contains purely those spin fluctuations longitudinal to the field. Figure~\ref{Szz}a shows the longitudinal excitation spectrum obtained from a data slice in the $[H~H~0]$ direction and Fig.~\ref{Szz}\textbf{b} the analogous MPS calculation. Both spectra show a weakly dispersive, low-energy branch running from X to M, and above this the entirety of continuum IV. To understand the origin of these longitudinal features, we appeal first to the Ising limit, where in the UUD phase a single spin-flip against the field direction costs no energy, whereas the opposite flip costs $3J_1$. If both processes occur on neighbouring spins, the energy cost is only $2J_1$ (Fig.~\ref{Szz}c), forming a localized two-magnon bound state. The spectrum close to the Ising limit then contains a nearly flat bound-state mode at an energy of $2J_1$, which is clearly split off from a continuum of states that starts around $3J_1$ (Supplementary Note 7). In Fig.~\ref{Szz}d we show an MPS calculation at rather strong Ising anisotropy that nevertheless shows many properties of the Heisenberg case (Fig.~\ref{Szz}b), and in Supplementary Figure 16 we show a more complete interpolation. These results demonstrate clearly the evolution of the lowest localized modes into a split-off and weakly dispersive longitudinal two-magnon bound state, while the upper localized modes evolve into a scattering resonance that forms the characteristic shape of continuum IV. Thus the Ising picture of these features remains valid even at the Heisenberg point.

Increasing the field into the V phase (Figs.~\ref{Dispersion}i-j) causes the longitudinal spectrum to show little change, whereas the transverse magnons disintegrate rapidly. Increasing non-collinearity leads to a mixing of transverse and longitudinal character, such that both sets of excitations merge into narrow continua (on the scale of the bandwidth) with strong intensity concentrated only at the $\Gamma$ and K points. These continuum features become both sharper and more dispersive with increasing field, regaining their single-magnon character above $B_{\mathrm{Sat}}$ (Fig.~\ref{camea}b). By contrast, as the field is decreased into the Y phase (Figs.~\ref{Dispersion}c-d), the effects of non-collinearity and dominant quantum fluctuations lead to a rapid loss of one-magnon character (again intensity is concentrated only at the $\Gamma$ and K points) and the emergence of wide excitation continua in both $S_{zz}$ and $S_{xx}$.

\bigskip
\noindent
{\bf DISCUSSION}
\smallskip

\noindent
Our INS measurements demonstrate unambiguously that the excitations of the TLHAF at all fields consist of magnon-like features only around the $\Gamma$ and K points that merge into extensive continua across the rest of the Brillouin zone. Despite the presence of at least short-ranged magnetic order at all fields, only in the UUD (1/3-plateau) phase can single magnons provide an adequate basis for describing the spectrum. Capturing the effects of strong quantum fluctuations on such weak order remains a major challenge, which we address by cylinder MPS calculations of the spectrum. Deploying such an unbiased numerical method allows one to divide the process of obtaining physical understanding into a two-step exercise of `expression' and `interpretation,' but brings into focus a dichotomy between the two. The expressibility of the MPS method is excellent, in that it captures all the features of the measured excitations with semi-quantitative accuracy, but as a numerical experiment its interpretability is limited. The primary contribution of our study is at the first step, namely providing an unbiased approach that confirms the true spectral content of a paradigm model. At least for the TLHAF, several different biased methods exist that interpret some of the observed spectral features, but to date have lacked a benchmark. Here it is the agreement between our INS and MPS results which allows us to assert that we have delivered the required benchmark.

We have in addition provided a modern standard for theoretical methods by employing the applied field as a control parameter to access four different, but continuously connected, physical regimes. Thus the ability to separate the transverse and longitudinal spectral functions in one regime affords some key insight that we use to interpret the longitudinal response in the other regimes. When we do consider one biased approach to interpreting the measured and calculated spectra, we find the hallmarks of bound and resonant states of magnon pairs. In the literature it has been argued that scattering continua can arise either from fractionalization (into bosonic \cite{Ghioldi2018,Zhang2022,Ghioldi2022} or fermionic \cite{Ferrari2019} components) or from the formation of two-particle and higher-order bound and scattering states of spin-1 excitations \cite{Zheng2006,Chernyshev2009,Powalski2015}, a subset of the latter being the magnon-breakdown scenario \cite{Stone2006,Zhitomirsky2013}. Although none of our present results necessitate a fractionalization scenario to explain the observed spectra, we certainly cannot exclude that fully quantitative analyses of the low-field limit could yet reveal the presence of deconfining $S = 1/2$ entities in the TLHAF.

To place our results in perspective, to our knowledge the complete field-induced spectrum of a 2D Heisenberg system has not previously been determined in experiment, and here we provide it for the TLHAF realized in CsYbSe$_2$. Methodologically, we have used our spectral data to benchmark cylinder MPS calculations of the dynamical spectral function at all applied fields, demonstrating that these now provide a powerful numerical method delivering near-quantitative accuracy. The next-neighbour TLHAF with $J_2$ on the cusp of the QSL phase provides a microcosm of all the key questions in quantum magnetism, arising where strong quantum spin fluctuations cause a partial or total suppression of magnetic order, whose extent can be controlled by an applied field. We believe that the combination of the three themes of our study, namely neutron spectroscopy in quantum materials, magnetic field-induced phenomena and MPS methods of accessing the complete spectral response of arbitrary locally interacting spin models, offers an exciting near-term future for quantum magnetism.

\bigskip
\noindent
{\bf METHODS}
\smallskip

\noindent
{\bf Experimental information}
\smallskip

\noindent
High-quality single crystals of CsYbSe$_2$ were prepared using the flux method \cite{Xing20191}. Refinement of single-crystal X-ray diffraction data demonstrated the complete absence of Cs/Yb site mixing, as detailed in Supplementary Note 1A. For the characterization of our crystals, we measured the magnetization up to 60~T in pulsed magnetic fields at the National High Magnetic Field Laboratory (MagLab) and the specific heat in a dilution refrigerator at temperatures down to 0.05~K and magnetic fields up to 9~T (results shown in Supplementary Notes 1C and 1D). Our electron spin resonance measurements were performed using a continuous-wave ESR spectrometer, collecting data at X-band frequencies ($\nu = 9.4$ GHz) and at $T = 15$ K. The resonance signal was measured from the field-derivative, $dP/dB$ of the power, $P$, absorbed in a transverse microwave magnetic field and the spectra were fitted to a Lorentzian lineshape.

Approximately 200 single-crystalline pieces totalling around 0.5~g of material were co-aligned on copper plates to obtain a mosaic sample shown in Supplementary Figure 4. Our neutron scattering experiments were performed on the time-of-flight (ToF) Cold Neutron Chopper Spectrometer (CNCS)~\cite{CNCS2} at the Spallation Neutron Source at Oak Ridge National Laboratory (ORNL), the multiplexing Continuous Angle Multiple Energy Analysis spectrometer (CAMEA)~\cite{CAMEA2}, and the cold-neutron Triple-Axis Spectrometer (TASP), the latter both located at the Swiss Spallation Neutron Source (SINQ) at the Paul Scherrer Institut (PSI). The measurements at CNCS were performed with an incident neutron energy $E_{\mathrm{i}} = 3.32$ meV, providing an energy resolution of 0.11~meV. A cryomagnet equipped with a dilution refrigerator was used to provide a maximum magnetic field of $B = 5$~T at temperatures down to 0.07 K. Measurements at CAMEA were performed with incident neutron energies $E_{\mathrm{i}} = 5.2$ and 6.2 meV (giving an energy resolution of 0.18~meV) and those at TASP with fixed $k_{\mathrm{i}} = k_{\mathrm{f}} = 1.5$~{\AA}$^{-1}$, both using an 11~T cryomagnet reaching a base temperature of $T \simeq\ 0.02$~K. In all three experiments, the sample was orientated in the $(HHL)$ scattering plane, such that the vertical magnetic field was applied along the [$-1$~1~0] direction in the $ab$ plane. The software packages~\textsc{MantidPlot}~\cite{Mantid} and~\textsc{Horace}~\cite{Horace} were employed for the data reduction and analysis at CNCS, while the data collected at CAMEA were analysed with the \textsc{MJOLNIR} software package~\cite{MJOLNIR}.

\medskip
\noindent
{\bf MPS calculations}
\smallskip

\noindent
We applied a cylinder MPS method to compute the dynamical spectral function of the isotropic spin-1/2 TLHAF in a magnetic field, as defined in Eq.~\eqref{Eq:Ham}, with $J_2 / J_1 = 0.03$. The MPS method proceeds by computing the time-dependent spin-spin correlation function
\begin{equation}
C^{\alpha \beta}_\mathbf{r} (\mathbf{x}, t) = \braket{\hat{S}^\alpha_{\mathbf{r} + \mathbf{x}}(t) \hat{S}^\beta_\mathbf{r}(0)},
\end{equation}
where $\mathbf{r}$ is the site at which the initial spin operator is applied, $\mathbf{x}$ is the vector separation in the two-point correlator, and $\alpha,\beta \in \{x, y, z\}$. The cylinder size, the bond dimension of the matrices used in the representation and the time-evolution procedures required to obtain well converged spectral functions at all fields are discussed and benchmarked in Supplementary Note 5. The calculations were implemented in Python using the package \textsc{TenPy} \cite{hauschild2018tenpy}.

The dynamical spin spectral function was obtained from the Fourier transform
\begin{equation}
\label{eq:cft}
S_{\mathbf{r}, \alpha \beta}(\mathbf{Q}, \omega) = \int_{-\infty}^{\infty} \mathrm{d}t \sum_\mathbf{x} e^{i(\omega t - \mathbf{Q} \cdot \mathbf{x})} C^{\alpha \beta}_\mathbf{r}(\mathbf{x}, t).
\end{equation}
The subscript $\mathbf{r}$ is retained because the correlation function is computed on a finite cylinder, with respect to the site $\mathbf{r}$ at its centre, and by time-evolving a ground state that breaks the translational symmetry of the Heisenberg model. To restore this symmetry in the spectral function, we average over three distinct time-evolved states, each corresponding to a site in the central unit cell, as explained in Supplementary Note 5. This procedure offers a strong reduction of the computational cost when compared with the MPS calculation of time evolution for a single spatially symmetric state. To account for artifacts in the spectral function caused by the finite cylinder length and time series in the Fourier transform, we convolve $C^{\alpha \beta}_\mathbf{r}(\mathbf{x}, t)$ in Eq.~\eqref{eq:cft} with a Gaussian envelope, as described in Supplementary Note 5. This results in an effective energy resolution of $0.1 J_1 \equiv 0.038$ meV and a momentum resolution of $0.032 / a \equiv 0.006$ r.l.u.

For comparison with the measured INS data, the calculated components of the dynamical structure factor were converted into a cross-section using the relation
\begin{eqnarray}
\hspace{-16pt} \frac{d^2\sigma}{d\Omega d\omega} \! \propto\ \! |F(\mathbf{Q})|^2 \! \sum_{\alpha,\beta} \! \left( \! \delta_{\alpha\beta} - \frac{Q_{\alpha}Q_{\beta}}{Q^2} \! \right) \!  (g_{\alpha\beta})^2 S_{\alpha\beta}(\mathbf{Q},\omega),
\label{Eq:cross-section}
\end{eqnarray}
where $F(\mathbf{Q})$ is the magnetic form factor of the Yb$^{3+}$ ion, $(\delta_{\alpha\beta} - Q_{\alpha} Q_{\beta}/Q^2)$ is the neutron scattering polarization factor and $g_{\alpha\beta}$ specifies the components of the $g$-tensor determined by ESR. A detailed comparison of our INS and MPS results is shown in Supplementary Note 6.

\bigskip
\noindent
{\bf DATA AVAILABILITY}
\smallskip

\noindent
The data that support the findings of this study are available from the corresponding author upon reasonable request.

\bigskip
\noindent
{\bf CODE AVAILABILITY}
\smallskip

\noindent
The code that supports the findings of this study is available from the corresponding author upon reasonable request.

\bigskip
\noindent
{\bf ACKNOWLEDGEMENTS}
\smallskip

\noindent
We thank C. McMillen for assistance with single-crystal X-ray diffraction refinements, J. Keum for assistance with X-ray Laue measurements, U. Nitzsche for technical assistance and B. Schmidt, A. S. Sukhanov and A. Chernyshev for helpful discussions. We acknowledge financial support from the Swiss National Science Foundation, from the European Research Council under the grant Hyper Quantum Criticality (HyperQC), the German Research Foundation (DFG) through the Collaborative Research Center SFB 1143 (project \# 247310070), the Austrian Science Fund FWF under project I-4548 and from the European Union Horizon 2020 research and innovation program under Marie Sk\l{}odowska-Curie Grant No.~884104. Research at Oak Ridge National Laboratory (ORNL) is supported by the U.S.~Department of Energy (DOE), Office of Science, Basic Energy Sciences (BES), Materials Sciences and Engineering Division. This research used the resources of the Spallation Neutron Source, a DOE Office of Science User Facility operated by ORNL. X-ray Laue alignment was conducted at the Center for Nanophase Materials Sciences (CNMS) (CNMS2019-R18) at ORNL, which is a DOE Office of Science User Facility. The pulsed magnetic field magnetometry measurements at National High Magnetic Field Laboratory are supported by the U.S. DOE, Office of Science, via BES program ``Science of 100 Tesla".

\bigskip
\noindent
{\bf AUTHOR CONTRIBUTIONS}
\smallskip

\noindent
Neutron scattering experiments and data refinement were performed by T.X., A.P. and S.E.N.
D.G.M., A.A.T. and P.G.N. assisted in performing the INS experiments at PSI.
Single-crystal growth and characterization were performed by J.X., L.D.S. and A.S.S.
Specific-heat measurements were performed by M.B. and P.K.
ESR measurements were performed by J.S.
Magnetization measurements were performed by J.X. and N.H.
Magnetization calculations were performed by S.N. and linear SWT calculations were performed by T.X. and S.E.N.
MPS calculations of spectral functions were performed by A.A.E. and A.M.L.
The manuscript was written by T.X., S.E.N., A.M.L. and B.N. with assistance from all the authors.

\bigskip
\noindent
{\bf COMPETING INTERESTS}
\smallskip

\noindent
The authors declare no competing interests.

\bigskip
\noindent
{\bf ADDITIONAL INFORMATION}
\smallskip

\noindent
{\bf Supplementary Information}
\smallskip

\noindent
The online version contains supplementary material available at
https://doi.org/xxx.yyy.zzz.

\bigskip
\noindent
{\bf Correspondence} and requests for materials should be addressed to Tao Xie, A.~M.~L\"auchli or S.~E.~Nikitin.

\end{document}


\onecolumngrid

\centerline{\large{\bf {Supplementary Information to accompany the article}}}

\vskip1mm

\centerline{\large{\bf {``Complete field-induced spectral response of the spin-1/2 triangular-lattice antiferromagnet CsYbSe$_2$"}}}

\vskip4mm

\centerline{Tao Xie, A. A. Eberharter, Jie Xing, S. Nishimoto, M. Brando, P. Khaneko, J. Sichelschmidt, A. A. Turrini, D. G. Mazzone, }

\vskip1mm

\centerline{P. G. Naumov, L. D. Sanjeewa, N. Harrison, Athena S. Sefat, B. Normand, A. M. L\"auchli, A. Podlesnyak and S. E. Nikitin}

\vskip8mm

\twocolumngrid

\section{Characterization}

\subsection{Single-Crystal X-Ray Diffraction}\label{sec:x-ray}

\begin{figure}[h]
\center{\includegraphics[width=\columnwidth]{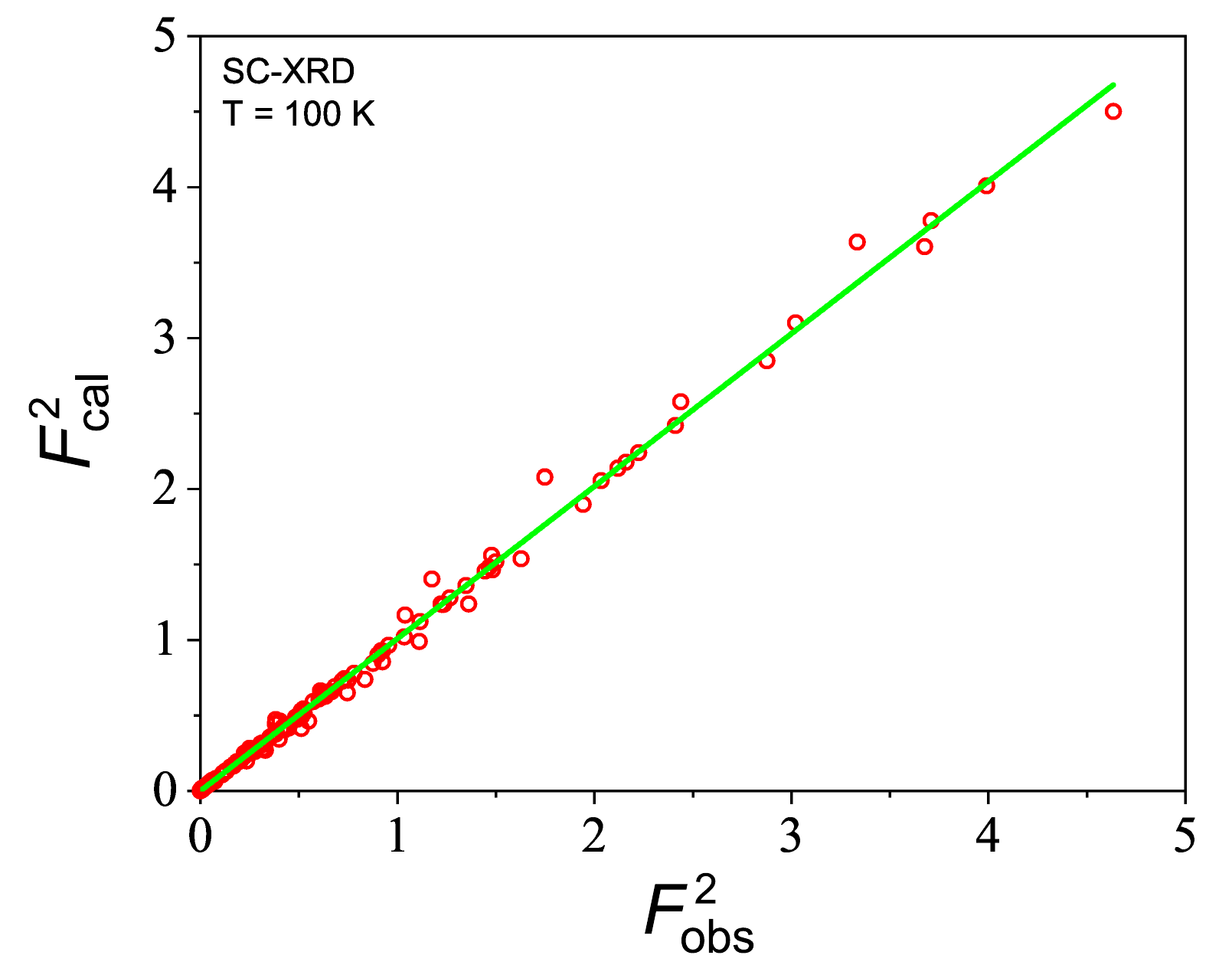}}
\caption{Rietveld refinement of single-crystal X-ray diffraction data at 100~K. $F^2_{\mathrm{obs}}$ and $F^2_{\mathrm{cal}}$ denote respectively the observed and calculated structure factors.}
\label{SCXRD}
\end{figure}

We have performed a careful investigation of the lattice structure of our CsYbSe$_2$ samples by analyzing several batches of single crystals using a Bruker Quest D8 single-crystal X-ray diffractometer. The structure was refined by the Rietveld method using the FullProf software package~\cite{FullProf}, which delivered excellent structural solutions with no evidence for site mixing. In these refinements, we treated the anisotropic displacement parameters and site occupancies as free variables, and in this way confirmed the complete absence of site disorder. The crystallographic data are presented in \ref{tabSCXRD}, and the result of a Rietveld refinement is shown in \ref{SCXRD}; we have also reported this crystal structure in the Cambridge Crystallographic Data Centre (CCDC)~\cite{CCDC1952075}. In addition we found no significant residual electron density, which would have suggested the presence of interstitial atoms. All of these results confirmed the high quality of our single crystals.

\begin{table}[b]
\caption{Crystallographic data for CsYbSe$_2$ determined by single-crystal X-ray diffraction.}
\begin{tabular}{|l|c|}
\hline
Empirical Formula                           & CsYbSe$_2$           \\ \hline
Formula weight (g/mol)                      & 463.87               \\ \hline
$T$, K                                      & 100                   \\ \hline
Crystal habit                               & red plates         \\ \hline
Crystal dimensions, mm                      & $0.14\times0.10\times0.03$ \\ \hline
Crystal system                              & hexagonal          \\ \hline
Space group                                 & P6$_3$/mmc (No.~194)   \\ \hline
\emph{a}, \AA                               & 4.1466(2)          \\ \hline
\emph{c}, \AA                               & 16.5050(1)         \\ \hline
Volume, \AA$^3$                             & 245.77(3)          \\ \hline
Z                                          & 2                  \\ \hline
Density (calc), g/cm$^3$                   & 6.268              \\ \hline
$\mu(\mathrm{Mo K_\alpha})$, mm$^{-1}$      & 40.932             \\ \hline
\emph{F}(000)                               & 386                \\ \hline
$T_{\rm{max}}$, $T_{\rm{min}}$                & 0.4415, 1.0000     \\ \hline
$\theta$ range for data collection          & 2.47-29.93         \\ \hline
Reflections collected                       & 3895               \\ \hline
Final $R~{[}I\textgreater   2\sigma(I){]} R_1, R_{w2}$ & 0.0274/0.0833      \\ \hline
Final $R$ (all data) $R_1, R_{w2}$          & 0.0284/0.0843      \\ \hline
Goodness of fit, $F^2$                    & 1.343              \\ \hline
\end{tabular}
\label{tabSCXRD}
\end{table}

\subsection{Crystal Structure and Next-Neighbour Magnetic Interaction}

One of the key questions in the analysis of every compound in the Yb-delafossite family is whether the ground state of the $S = 1/2$ system can be a quantum spin liquid (QSL) at zero applied magnetic field. In our work we have confirmed the presence of 120$^{\circ}$ order in CsYbSe$_2$, at temperatures below 0.4~K and at least over a spatial range far exceeding the lattice constant [Fig.~1c of the main text and \ref{correlationlength}(e) below]. The same order is found in KYbSe$_2$~\cite{scheie2021}, but notably not in NaYbSe$_2$~\cite{dai2020spinon}. Factors destabilizing this order include a possible frustrated coupling between triangular-lattice (TL) planes and a possible next-neighbour interaction within the planes, where (as noted in the main text) recent numerical studies have achieved partial agreement that the ground state is a QSL in the regime $0.06 \lesssim J_2/J_1 \lesssim 0.15$~\cite{Kaneko2014,Li2015,Zhu2015,Hu2015,Iqbal2016,Hu2019}.

\begin{table*}[tb]
\caption{Crystallographic information for selected Yb-selenide delafossites~\cite{xing2021synthesis}. }
\begin{tabular}{|l||c|c|c|c|}
\hline
$\;$ Chemical formula     & NaYbSe$_2$        & KYbSe$_2$         & RbYbSe$_2$            & CsYbSe$_2$            \\ \hline $\;$ Crystal system       & trigonal          & trigonal          & trigonal              & hexagonal             \\ \hline
$\;$ Space group          & $\;$ R$\overline{3}m$ (No.~166) $\;$ & $\;$ R$\overline{3}m$ (No.~166) $\;$ & $\;$ R$\overline{3}m$ (No.~166) $\;$ & $\;$ $P6_3/mmc$ (No.~194) $\;$ \\ \hline
$\;$ Stacking type        & ABC               & ABC               & ABC                    & AA                    \\ \hline
$\;$ Ionic radius, \AA     & 1.02 (Na$^+$)     & 1.38 (K$^+$)      & 1.52 (Rb$^+$)         & 1.67 (Cs$^+$)         \\ \hline
$\;$ Interlayer distance, \AA $\;$ &  6.92       & 7.56              & 7.88                  & 8.25                  \\ \hline
\end{tabular}
\label{delafossite}
\end{table*}

Addressing first the issue of layer stacking, the majority of Yb delafossites have a structure described by the R$\overline{3}m$ space group, as listed in \ref{delafossite}, which has an ABC stacking that does indeed suggest frustration of antiferromagnetic (AF) interlayer couplings. However, some members of the series with larger alkali-metal ions, including Cs (\ref{delafossite}), have a structure with space group P$6_3/mmc$ that has AA stacking, and hence no interlayer frustration. Thus the available materials examples tend to suggest that the stacking of TL layers is not a relevant factor. Turning to $J_2$, in our work we have deduced a weak next-neighbour interaction in CsYbSe$_2$, $J_2 = 0.03 J_1$. The authors of Ref.~\cite{scheie2021} deduced a value $J_2 = 0.05 J_1$ in KYbSe$_2$, placing it closer to the QSL regime. While the authors of Ref.~\cite{dai2020spinon} were not able to determine a $J_2$ value, one may certainly suggest that the larger lattice constant inherent to the members with larger alkali-metal ions (\ref{delafossite}) causes a reduction of $J_2$ and hence an increasing stabilization of the 120$^\circ$-ordered ground state at base temperature in the Yb-delafossite materials.

\begin{figure}[t]
\center{\includegraphics[width=0.8\columnwidth]{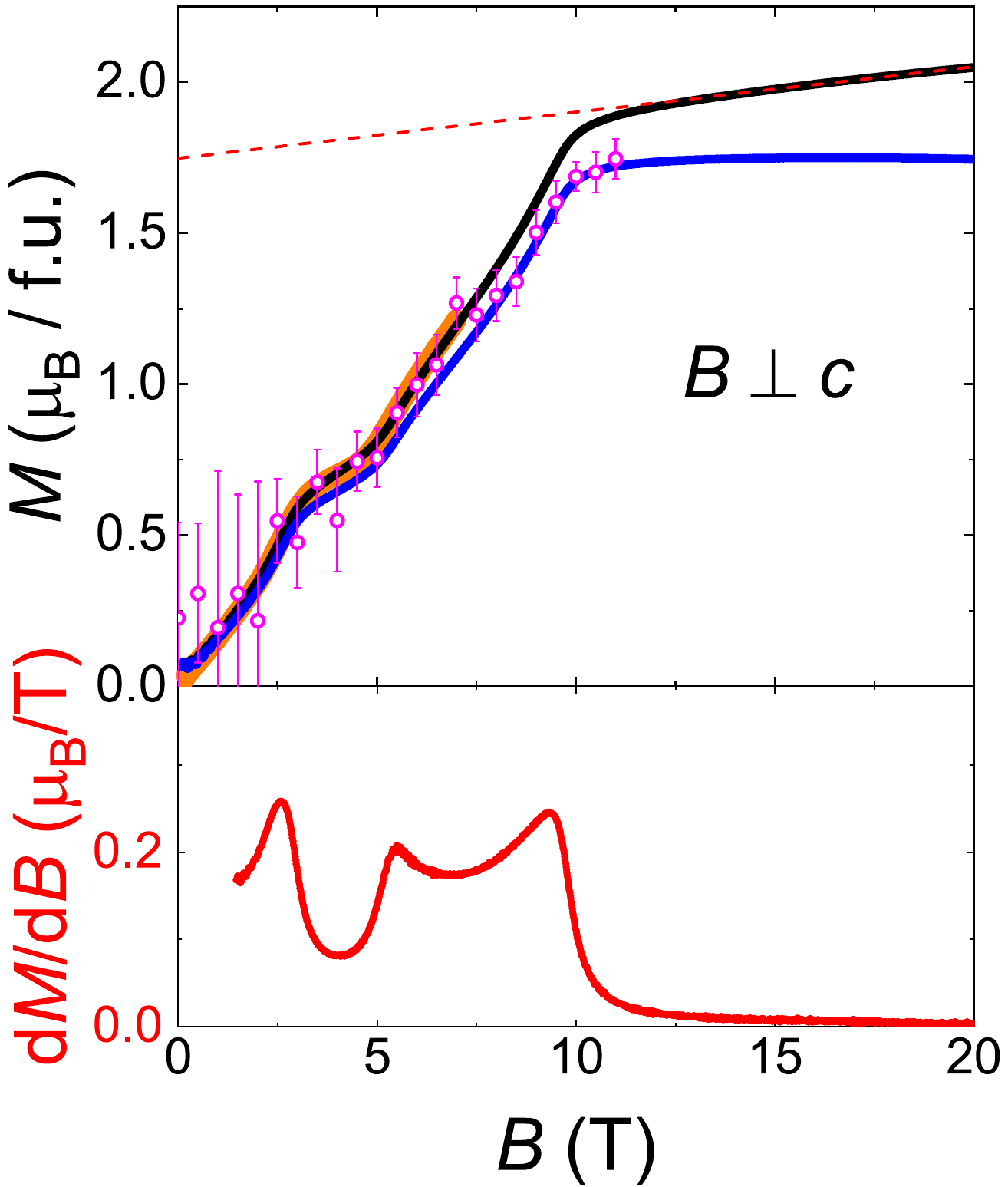}}
\caption{Isothermal magnetization measured in a pulsed magnetic field. The black curve shows the raw magnetization data, the red dashed line the estimated van Vleck contribution and the blue curve the intrinsic magnetization obtained by subtraction of the van Vleck part. The orange symbols are the low-field (up to 7 T) magnetization data measured in a MPMS-7. Open circles show the bulk moment extracted from the field-dependence of the (0,~0,~4) Bragg peak measured by neutron diffraction at $T < 0.05$~K. The solid red curve in the lower panel shows the first derivative of the magnetization, $dM/dB$.}
\label{MHhigh}
\end{figure}

\begin{figure}[t]
\center{\includegraphics[width=0.75\columnwidth]{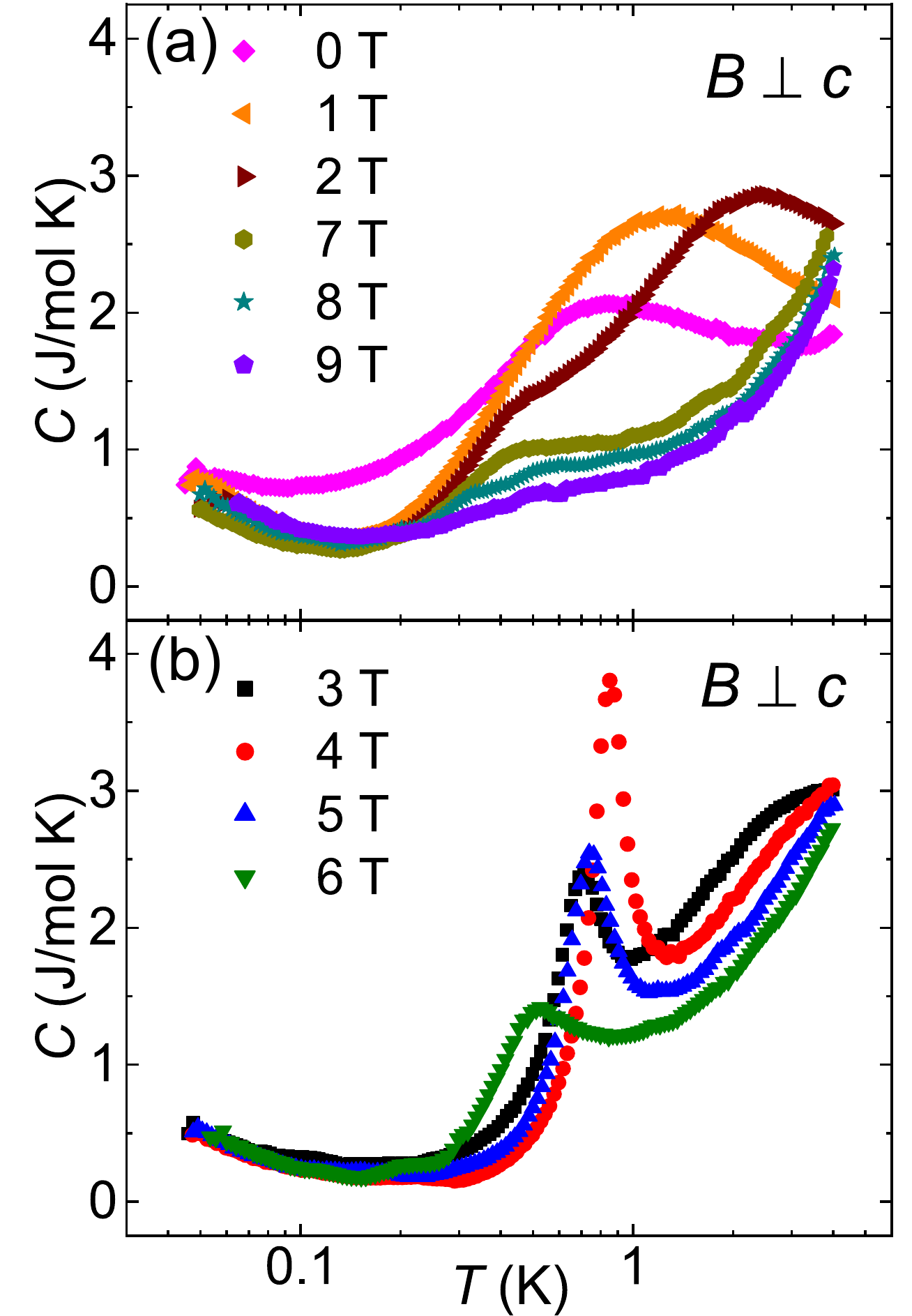}}
\caption{Temperature-dependence of the specific heat measured in a range of applied in-plane magnetic fields ($\mathbf{B} \perp {\hat c}$).}
\label{heat}
\end{figure}

\begin{figure*}[t]
\center{\includegraphics[width=0.82\textwidth]{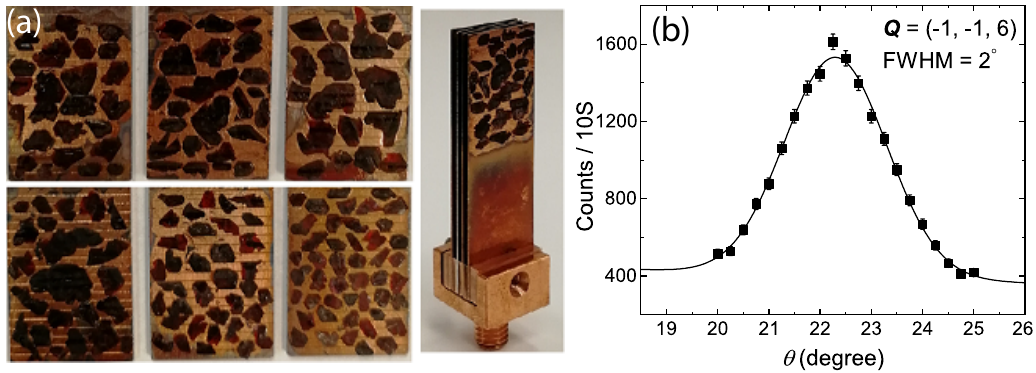}}
\caption{(a) Photographs of the CsYbSe$_2$ crystals coaligned and assembled for neutron scattering experiments. (b) Rocking curve of the coaligned sample measured by neutron diffraction, with a Gaussian fit shown by the solid line.}
\label{sample}
\end{figure*}

\subsection{Magnetization in Pulsed Magnetic Field}

We measured the isothermal magnetization at $T = 0.4$~K in pulsed magnetic fields up to 60~T, and for calibration in a MPMS-7 with a $^3$He insert at fields up to 7~T. We note that the magnetic field in all our experiments is applied in the $ab$ plane, unless otherwise stated, and that we do not distinguish between in-plane directions. In the raw magnetization data shown by the black solid line in \ref{MHhigh}, we observe a clear and continuous increase above a saturation field of approximately 10~T. This contribution results from van Vleck paramagnetism and can be approximated in lowest order by a linear function (the dashed red line), from which the van Vleck susceptibility may be estimated as $\chi^{VV} \approx 0.0152~\mu_B$/T. Subtracting this contribution from the raw data~\cite{ranjith2019field,Ranjith2019naybse,ding2019gapless} leaves an intrinsic magnetization [solid blue line in \ref{MHhigh}, shown also in Fig.~1c of the main text] displaying near-perfect saturation above a field $B_{\mathrm{Sat}}$ that we estimate most accurately from the excitation spectrum of the fully polarized phase [Fig.~2b of the main text]. In \ref{MHhigh} we show also the bulk magnetization obtained from the field-induced enhancement of the (0~0~4) Bragg peak measured on TASP at the Paul Scherrer Institute (PSI), and these results agree quantitatively with the direct measurements. We defer the accurate modelling of these data, which we performed to obtain the orange line in Fig.~1c of the main text, to \ref{sec:mag}.

\subsection{Specific Heat}

We measured the specific heat in a dilution refrigerator at temperatures down to 50 mK and magnetic fields up to 9 T. A sharp peak in $C(T)$ can be used to establish the presence of long-range order in the system. In \ref{heat}(a) we group the applied fields, namely those below 3~T and above 6~T, in which no clear peak appears in $C(T)$ at all, only a broad hump at the low fields and a weak shoulder at the high fields. In \ref{heat}(b) we show that relatively sharp, $\lambda$-shaped peaks can be found at 3, 4 and 5~T, in partial correspondence with the long-range order of the UUD phase found by neutron diffraction. The temperatures of these sharp peaks are indicated as solid circles in the phase diagram shown in Fig.~1d of the main text and the temperatures of the features found at $B \le 2$~T and $B \ge 6$~T as open circles.

In more detail, the specific heat at zero field in \ref{heat}(a) shows a broad hump with no trace of a phase transition. This should be contrasted with the neutron diffraction measurements in Fig.~1b of the main text, which display a series of weak magnetic Bragg peaks with a clear onset at $T \simeq 0.4$~K. However, the analysis of the correlation lengths presented in \ref{sec:neutron_scattering}B leads to the result that these are finite at zero field, signalling a lack of true, long-ranged AF order even at $T = 0.02$~K, which is consistent with the absence of a sharp peak in the corresponding $C(T)$ data. The specific heat does show sharp, $\lambda$-shaped peaks in the field range 3-5~T [\ref{heat}(b)] that reflect phase transitions into a long-range-ordered phase on, and apparently near, the 1/3-plateau state. These thermodynamic data are confirmed by the field-dependence of the (1/3, 1/3, 1) magnetic peak presented in Fig.~1c of the main text and analyzed in detail in \ref{fig_tasp} below.

\begin{figure*}[t]
\center{\includegraphics[width=0.9\textwidth]{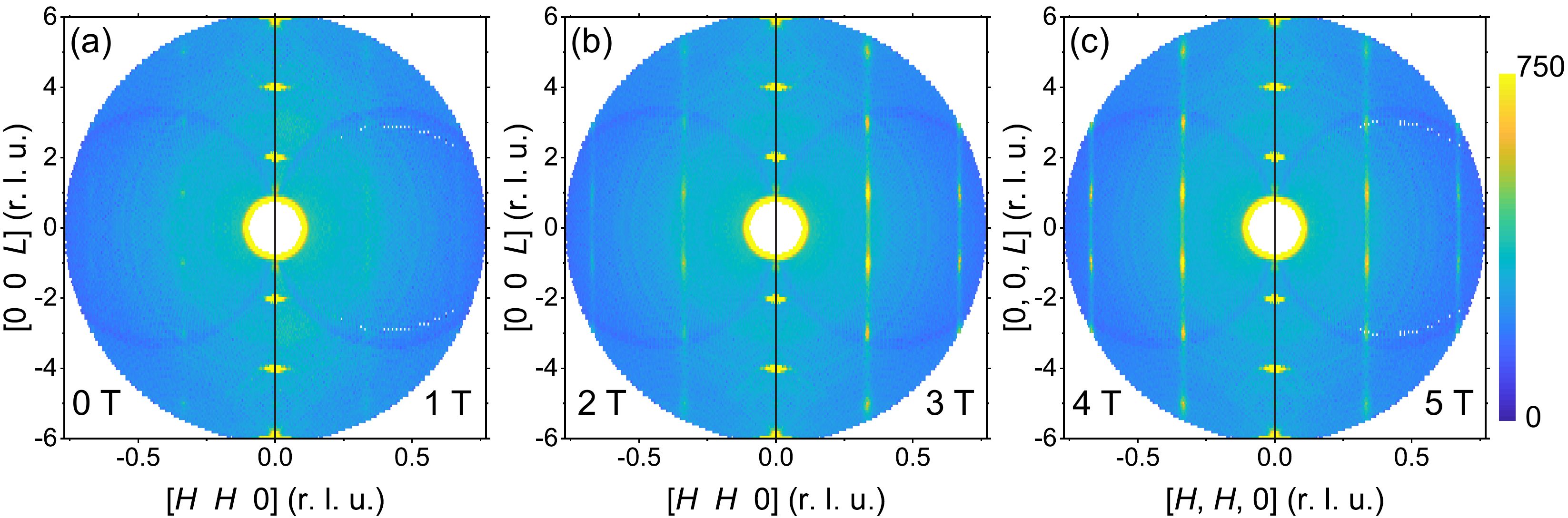}}
\caption{Constant-energy slices in the $(H~H~L)$ plane taken at zero energy transfer under different magnetic fields, with the data symmetrized about the $[H~H~0]$ axis. The strong spots at $Q = (0,~0,~L)$ for even-integer $L$ are nuclear Bragg peaks.}
\label{2D_elastic}
\end{figure*}

\begin{figure*}[t]
\center{\includegraphics[width=1\textwidth]{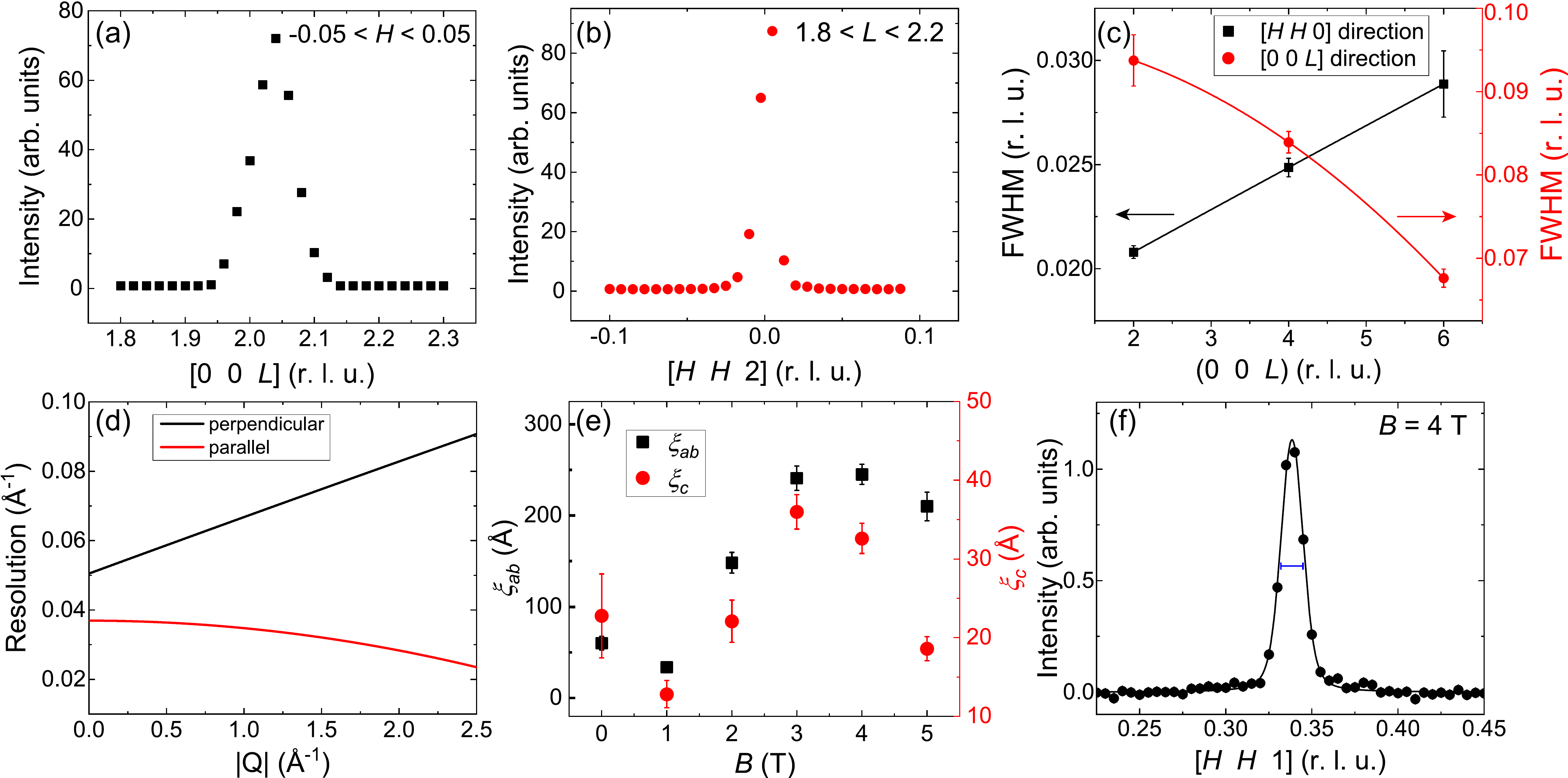}}
\caption{(a,b) 1D intensity cuts at zero energy along the $[0~0~L]$ and $[H~H~2]$ directions at the (0, 0, 2) nuclear Bragg peak at 0 T. (c) FWHM of nuclear peaks obtained from cuts along the [$H~H$ 0] direction (black) and $[0~0~L]$ direction (red) at $Q$ = (0, 0, 2), (0, 0, 4) and (0, 0, 6). Solid lines in both panels show polynomial fits. (d) Perpendicular and parallel $|Q|$-resolution estimated from the FWHMs shown in panel (c). (e) Perpendicular ($\xi_{\perp}$) and parallel ($\xi_{\parallel}$) correlation lengths calculated from the magnetic Bragg peaks shown in \ref{2D_elastic}. For this geometry $\xi_{\perp} \equiv \xi_{ab}$, the in-plane correlation length, and $\xi_{\parallel} \equiv \xi_c$, the out-of-plane correlation length. (f) An example 1D cut along the [$H~H~1$] direction through the magnetic Bragg peak (1/3,~1/3,~1) at 4 T. The horizontal blue bar represents the instrumental $|Q|$-resolution, which indicates a resolution-limited peak at 4 T.}
\label{correlationlength}
\end{figure*}

\section{Neutron Scattering Experiments}
\label{sec:neutron_scattering}

\subsection{Sample Preparation}

We coaligned around 200 single crystallites on copper plates, as shown in \ref{sample}(a), to obtain a mosaic sample with a total mass of approximately 0.5 g. The rocking curve obtained by a neutron diffraction measurement of one selected magnetic peak [\ref{sample}(b)] shows that the mosaicity of this sample is approximately $2\degree$ FWHM.

\subsection{Elastic Neutron Scattering at CNCS}
\label{sec:xi}

\ref{2D_elastic} shows two-dimensional (2D) slices through the measured intensity dataset at zero energy in the ($H~H~L$) plane at fields from 0 to 5 T. At 0~T we observe weak but nonetheless clear intensity peaks at $Q = (1/3,~1/3,~L)$ for odd-integer $L$, which become weaker and almost invisible at 1 T. For $B \ge 2$ T we find rods of magnetic intensity elongated along the $[1/3~1/3~L]$ direction, with peaks at the same $Q$ values. The intensities of these peaks signal effectively long-ranged order around 3-5 T, their very broad nature in $L$ reflects the 2D nature of the magnetic system (as opposed to new peaks at intermediate $L$ values) and their $(H,~K)$ position confirms the persistence of threefold periodicity in the plane of the triangular lattice at all applied fields. In the context of \ref{2D_elastic}(a), we state for completeness that the data shown in the upper panel of Fig.~1b of the main text were integrated over the interval $K = [-0.05, 0.05]$, symmetrized according to the crystal symmetry and unfolded for visual clarity, meaning that the intensities at $\pm L$ are equivalent.

For a specific magnetic peak, the spin correlation length can be estimated using the formula $\xi = 2\pi/\sqrt{w^2 - R^2}$~\cite{Young2013}, where $w$ is the full width at half maximum (FWHM) height of the magnetic peak and $R$ is the instrumental momentum resolution at this peak. To estimate this resolution, we first prepared 1D intensity cuts along the [$H$~$H$~0] and [0~0~$L$] directions at the nuclear peaks (0,~0,~2) [\ref{correlationlength}(a,b)], (0,~0,~4) and (0,~0,~6), then fitted a Gaussian function to them. The FWHM values of these three nuclear peaks reflect the momentum resolution of the instrument perpendicular (for [$H$~$H$~0]) and parallel (for [0~0~$L$]) to ${\hat Q}$ at each reciprocal-lattice point. Polynomial fits to the FWHM as a function of $L$ [\ref{correlationlength}(c)] yield approximate estimates of the resolution as a function of $\left|\bf{Q}\right|$ for an arbitrarily chosen momentum transfer, $\mathbf{Q}$, in the perpendicular and parallel directions [\ref{correlationlength}(d)].

To obtain the FWHM of the magnetic peaks for a quantitative determination of the correlation length, we fitted 1D cuts through the magnetic peaks to the Voigt function, which is defined as
\begin{align}
y(x) & = y_{0} + (f_1 * f_2) (x),
\label{Voigt}
\end{align}
where the $*$ denotes a convolution. In Eq.~\eqref{Voigt}, $f_{1}(x) = \frac{2A}{\pi} \frac{w_{L}}{{4(x - x_{c}})^2 + w_{L}^2}$ is the Lorentz function, with peak centre $x_c$ and FWHM $w_{L}$, and $f_{2}(x) = \sqrt{\frac{4\ln 2}{\pi}} \frac{1}{w_{G}} \exp (-\frac{4\ln 2}{w_{G}^2} x^2)$ is a Gaussian with peak centre $x = 0$, unit peak area and FWHM $w_{G}$, which should be fixed to the instrumental resolution, $R(\left|\bf{Q}\right|)$, for a specific $\bf{Q}$. The FWHM of the Voigt function is then $w_V$ = $0.5346 w_{L} + \sqrt{0.2166\cdot w_{L}^2 + w_{G}^2}$. By inserting $w_V$ for $w$ and $R(\left|\bf{Q}\right|)$ for $R$ into the above expression for $\xi$, we estimate the correlation lengths at $\mathbf{Q}$ = (1/3, 1/3, 1) for each different field. \ref{correlationlength}(e) shows the correlation lengths we deduce from all of our CNCS data, which from our choice of the (0~0~$L$) series of nuclear peaks correspond to the in- and out-of-plane correlations. At $B = 0$ we obtain the values $\xi_{ab} = 60(7)$~{\AA} and $\xi_c = 23(5)$~{\AA} quoted in the main text. We observe that the correlation lengths dip at 1 T before $\xi_{ab}$ rises strongly, to values in excess of 200~{\AA}, on and just above the 1/3 plateau. To gauge the meaning of this number, in \ref{correlationlength}(f) we show a 1D cut along $[H~H~1]$ at $B = 4$ T: the horizontal blue bar is the $|Q|$-resolution at the magnetic Bragg peak (1/3, 1/3, 1), which is clearly resolution-limited, and thus the intrinsic $\xi$ values diverge as expected for the long-range-ordered UUD state.

\subsection{Elastic Neutron Scattering at TASP}
\label{sec:etasp}

\begin{figure}[t]
\center{\includegraphics[width=\columnwidth]{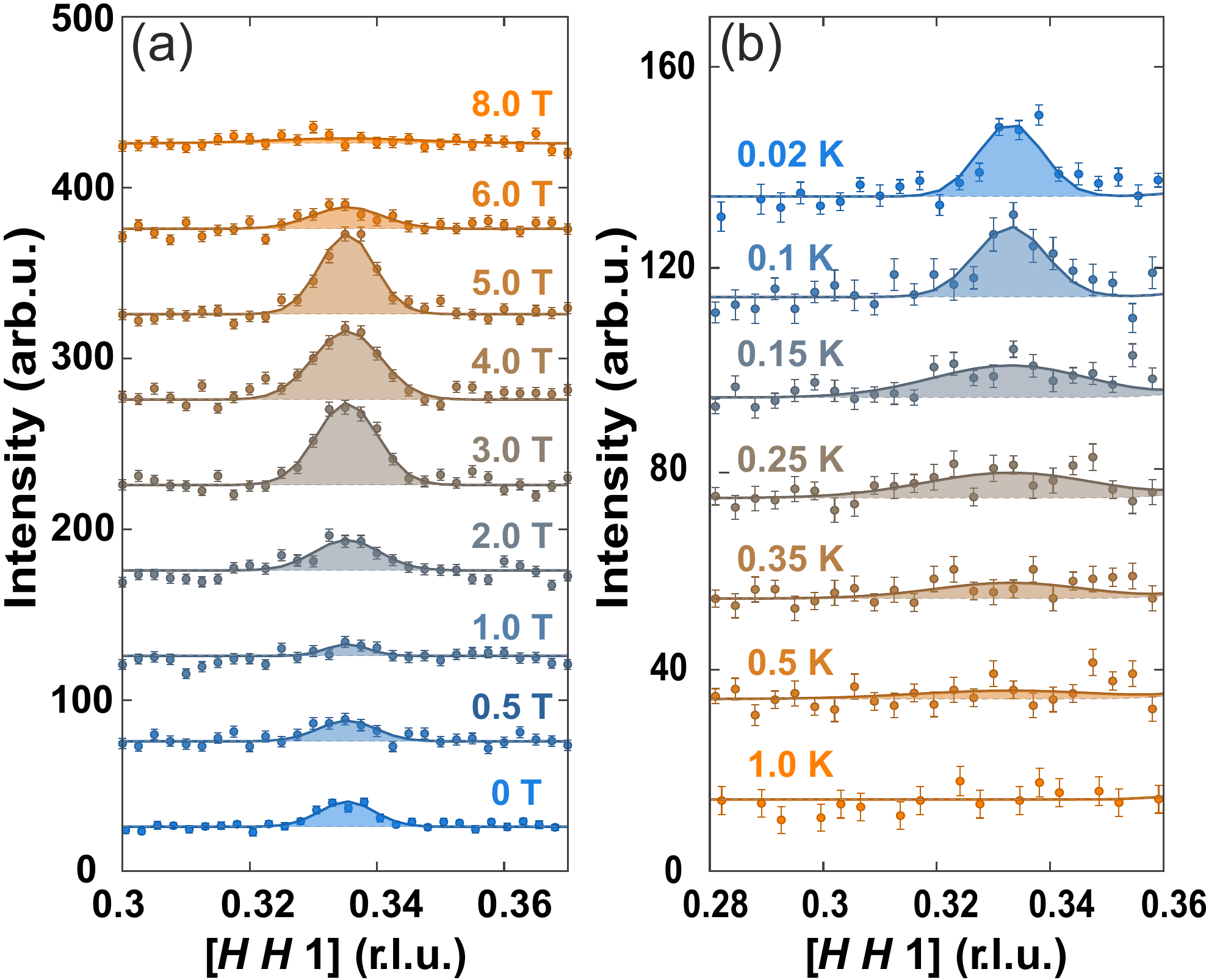}}
\caption{(a)~Intensity of the ($1/3$,~$1/3$,~1) magnetic Bragg peak measured at $T < 0.05$~K for a series of magnetic fields, which are offset by 50 units for clarity. Solid lines and shading show Gaussian fits. (b) Intensity of this peak measured at zero field for a series of temperatures below 1~K, which are offset by 20 units for clarity. }
\label{fig_tasp}
\end{figure}

The dependence of the integrated intensity of the magnetic peak (1/3,~1/3,~1) on the applied magnetic field is summarized in Fig.~1c of the main text and its dependence on temperature in Fig.~1b. \ref{fig_tasp}(a) shows the raw elastic neutron scattering data measured on TASP in scans along the $[H~H~1]$ direction at base temperature for a series of fields. The peak intensity is clearly suppressed by a weak field, with a minimum around 1~T, before becoming sharper and stronger as the field is increased, reaching a maximum around 4~T. This intensity then decreases beyond 5~T to very small values higher in the V regime. \ref{fig_tasp}(b) shows data obtained for the same scan at zero field for multiple temperatures below 1~K, where the robust peak present at $T = 0.35$ K clearly becomes sharper and stronger towards base temperature.

\subsection{Two-Dimensional Excitation Character}

\begin{figure*}[t]
\center{\includegraphics[width=1\textwidth]{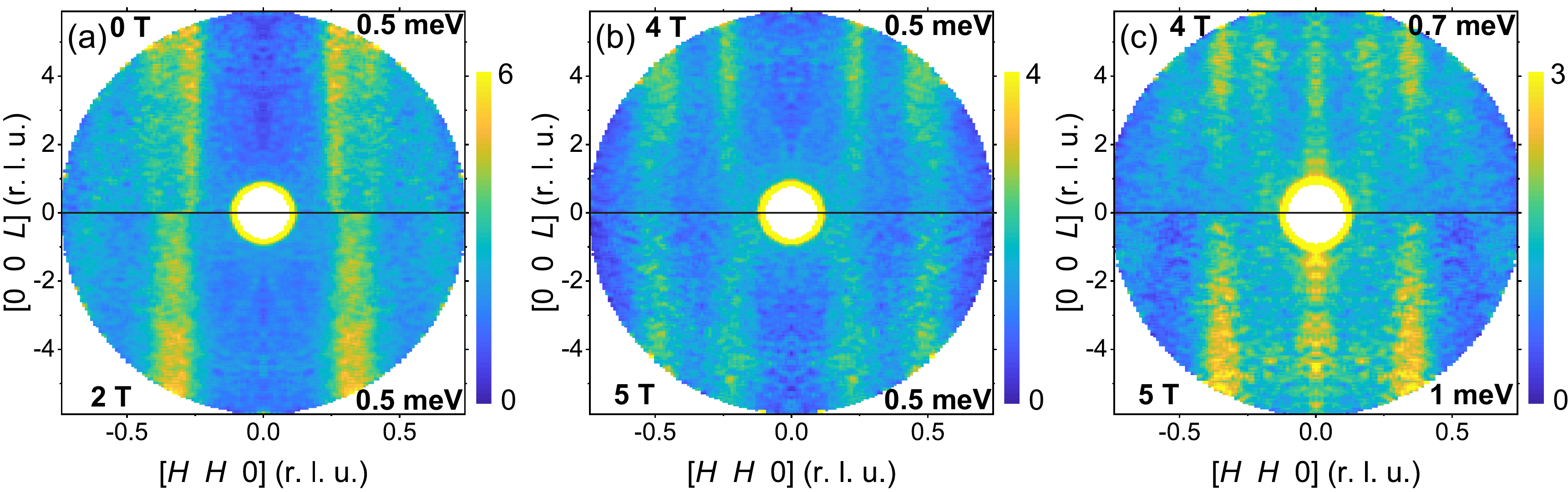}}
\caption{Constant-energy 2D intensity slices at different energies and magnetic fields. The data have been symmetrized about the [0~0~$L$] axis.}
\label{2D_inelastic}
\end{figure*}

The spin excitations we observe in our measurements retain their highly 2D nature over the full energy range and under all applied fields. In the representative 2D constant-energy slices presented in \ref{2D_inelastic}, the intensity distribution in all cases takes the form of multiple rods extending largely unchanged along the $L$ direction. This definitive proof of 2D spin excitations is fully consistent with previous reports on CsYbSe$_2$~\cite{Xing20192} and related delafossites~\cite{dai2020spinon,scheie2021}. As a consequence we may integrate our intensity data over a wide $L$ range to analyse the in-plane spin dynamics, and for the data shown both below and in Fig.~3 of the main text, this integration range, $1.2 \le L \le 3.8$, is denoted by the $L$ label 2.5.

\begin{figure*}[t]
\center{\includegraphics[width=0.8\textwidth]{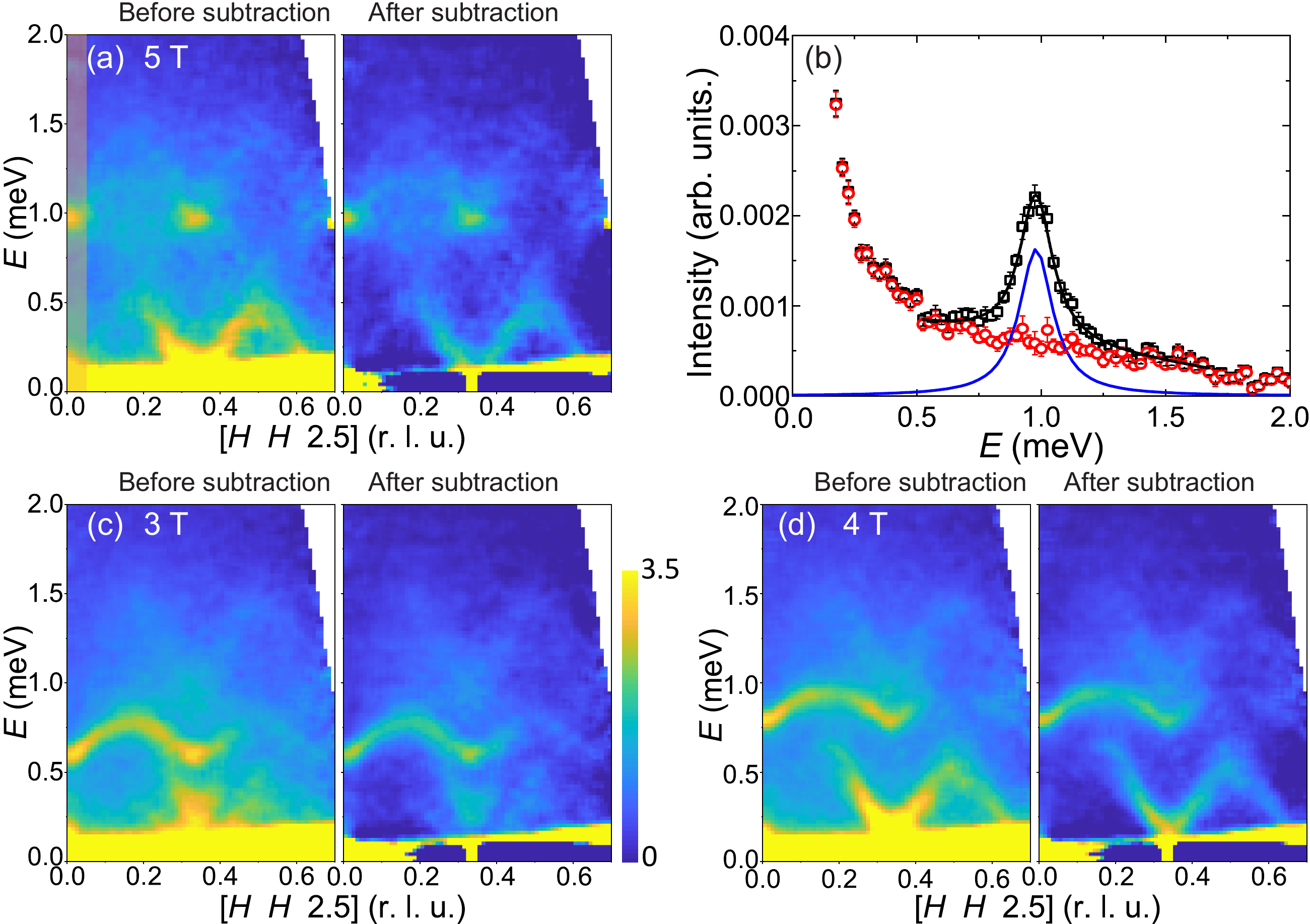}}
\caption{Definition of the subtracted background in the INS spectra and illustration of selected INS spectra before and after background subtraction. (a) Left: raw data for the spin excitation spectrum along $[H~H~2.5]$ at 5 T and 70 mK; right: corresponding spectrum obtained after background subtraction. This background is extracted from the shaded area ($0 \le H \le 0.05$). (b) Open black squares show the 1D energy cut obtained by integrating over the shaded region in panel (a). The solid black line shows a Lorentz fit to this cut with a linear background, the blue line shows the Lorentz peak fit alone, and the open red circles show the difference, which is used as the real background for subtraction purposes. (c) INS spectra before and after background subtraction for $B = 3$ T. (d) Spectra before and after subtraction for $B = 4$ T.}
\label{BKG}
\end{figure*}

\subsection{Background Definition and Subtraction for INS Spectra}

The INS spectrum is always contaminated by a background contribution that arises from incoherent neutron scattering and from scattering due to the sample environment. Raw spectral data obtained on CNCS for the $[H~H~2.5]$ direction (meaning with the broad $L$ integration described above) are shown in the left panels of \ref{BKG}(a,c,d) for three different magnetic fields at our base temperature of 70 mK. For an accurate characterization of the background, we make use of the fact that the 5 T spectrum at the $\Gamma$ point, $I(\mathbf{Q} = \mathbf{0}, E)$, consists of a single and well-defined inelastic peak near 1 meV. We assume that an appropriate 1D intensity cut, prepared from the shaded region in \ref{BKG}(a) and shown by the black points in \ref{BKG}(b), can be described within the energy window 0.5~meV~$\le E \le$~1.5~meV by a Lorentzian peak and a linear background. Thus we fitted the measured signal to the form
\begin{align}
I(E) = a_0 + a_1 E + \frac{I_0 W_0^2}{(E - E_0)^2 + W_0^2},
\label{eq:BG}
\end{align}
where $I_0$, $W_0$ and $E_0$ characterize respectively the intensity, width and centre of the inelastic peak. We subtract the fitted Lorentzian, shown by the blue line in \ref{BKG}(b), from the raw spectrum to obtain a residual intensity that we consider as fully representative of the background [shown by the red points in \ref{BKG}(b)]. The results of subtracting this background are shown in the right panel of \ref{BKG}(a), where clear limits to the extent of each continuum become visible. Because our data provide no evidence that the background varies with $\mathbf{Q}$, we subtracted the same form from our cuts everywhere in reciprocal space. Similarly, the background has no dependence on the magnetic field, and examples of full background-subtracted spectra at $B = 3$ and 4 T are shown in \ref{BKG}(c,d). We stress that this method provides a minimal parameter-free background model, and its simplicity far outweighs the disadvantage of a minor (statistically insignificant) oversubtraction appearing at some high $H$ values.

\begin{figure}[t]
\center{\includegraphics[width=\columnwidth]{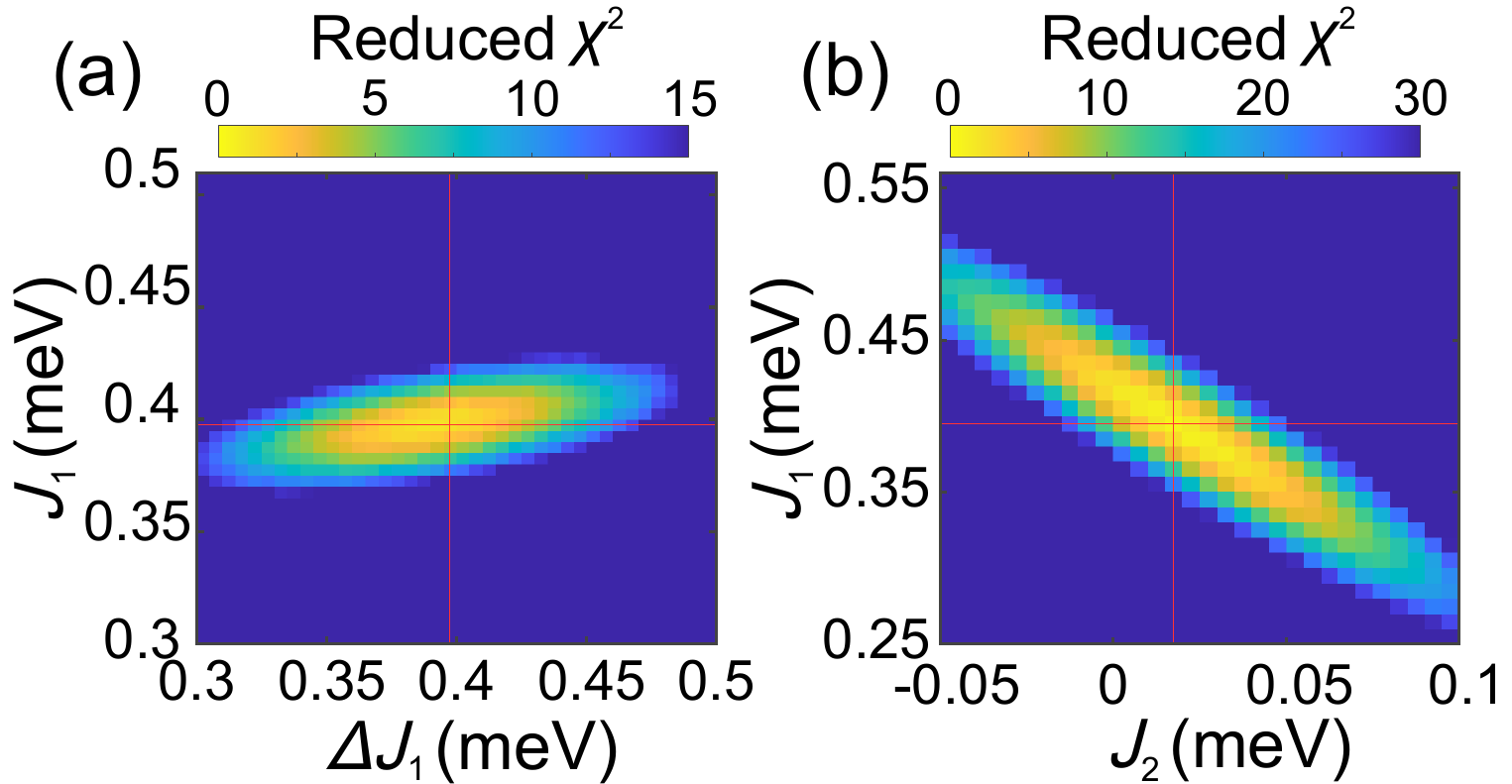}}
\caption{(a)~Fit quality shown as a function of assumed XXZ-model parameters $J_{1}$ and $\Delta$ for the dominant nearest-neighbour interaction. Here $J_2$ has been fixed to its optimal value of 0.011 meV. (b)~Fit quality shown as a function of $J_1$ and $J_2$, assuming $\Delta\ = 1$. Red crosses in both panels indicate the position of the global minimum.}
\label{Fig:fit_quality}
\end{figure}

\section{Linear Spin-Wave Theory}
\label{sec:lswt}

In our analysis we use linear spin-wave theory (SWT) to achieve two separate goals. Working above the saturation field, $B_{\mathrm{Sat}}$, we make use of the fact that the excitations are guaranteed to be well defined spin waves to obtain the most accurate available fit of the magnetic interaction parameters of CsYbSe$_2$ (\ref{sec:lswt}A). Working below $B_{\mathrm{Sat}}$, we use linear SWT for a preliminary indication of the locations and energy scales of putative $\Delta S = 1$ excitations in the TLHAF, and hence of departures from semiclassical magnetism arising due to quantum corrections. In \ref{sec:lswt}B we outline how the SWT results shown in Fig.~3 of the main text were obtained and use these to obtain an overview of possible two-magnon contributions.

\subsection{Fitting of Magnetic Interactions}

As described in the main text, we determine the magnetic interactions by using the 11~T dataset from CAMEA, which shows a single, sharp magnon mode with its maximum at the $\Gamma$ point [Fig.~2b of the main text]. We quantified the location of this mode at 17 points in reciprocal space and used $\textsc{SpinW}$ \cite{spinw} to fit these to the Hamiltonian
\begin{eqnarray}
\mathcal{H} & = & J_1 \sum_{\langle i,j \rangle} ({S}^x_i{S}^x_j + {S}^z_i{S}^z_j + \Delta{S}^y_i{S}^y_j ) \nonumber \\
& & + J_2 \sum_{\langle\langle i,j \rangle \rangle} \mathbf{S}_i\cdot\mathbf{S}_j - \mu_B g_{ab} B \sum_i{S}^z_i.
\label{Eq:Ham}
\end{eqnarray}
The optimal fit in the 3D parameter space of $J_{1}$, $\Delta$ and $J_2$ is $J_{1} = 0.395(7)$~meV, $\Delta J_{1} = 0.39(2)$~meV and $J_2$ = 0.011(4)~meV. In \ref{Fig:fit_quality} we illustrate the quality of this fit by showing two 2D cross-sections: fixing $J_2$ to its optimal value [\ref{Fig:fit_quality}(a)] shows that $J_{1} = \Delta J_{1}$, i.e.~the nearest-neighbour interaction is isotropic within the precision of our measurements; setting $\Delta = 1$ [\ref{Fig:fit_quality}(b)] allows us to optimize $J_2$, and hence to conclude that magnetic Hamiltonian of CsYbSe$_2$ is described by the $J_1$-$J_2$ Heisenberg model with $J_1 = 0.395(8)$~meV and $J_2 = 0.011(4)$~meV ($J_2/J_1$ = 0.03).

\begin{figure*}[t]
\centering
\includegraphics[width=\textwidth]{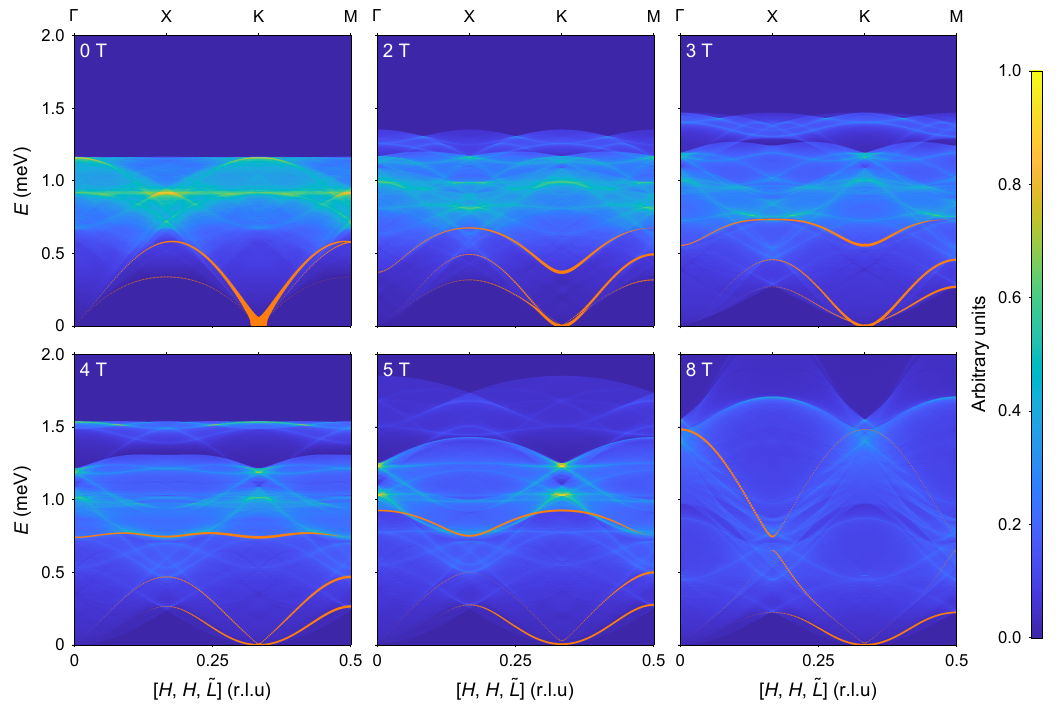}
\caption{Two-magnon densities of states computed from linear SWT at six different magnetic fields for the $J_1$-$J_2$ TLHAF with $J_2$/$J_1$ = 0.03. ${\tilde L}$ denotes integration over the full range of $L$, equivalent to an ideally two-dimensional system.}
\label{dos2m}
\end{figure*}

\subsection{SWT Spectra}

Linear SWT ceases to provide a complete description of the spectrum of the TLHAF as soon as the field is lowered below $B_{\mathrm{Sat}}$, when quantum corrections become finite. The TLHAF in an in-plane magnetic field has long been known~\cite{chubokov1991quantum} to exhibit a deformed 120$^{\circ}$ phase (Y), the 1/3 plateau phase (UUD) and a V (or 2:1) phase, with two of the spins parallel, below saturation, as illustrated schematically in Fig.~1d of the main text. At the mean-field level, the magnetization is perfectly linear in field from 0 to $B_{\mathrm{Sat}} = 9.6$~T (Fig.~\ref{DMRG_MH}), which is obtained exactly due to the classical nature of the problem at $B \geq B_{\mathrm{Sat}}$. However, $M(B)$ has no plateau around $B_{\mathrm{Sat}}/3$, as this phase is stabilized only at higher order in $1/S$. In linear SWT, there is a large family of classically degenerate states at any field $B < B_{\mathrm{Sat}}$, and the precise orientation of the Y, UUD or V state at any given field is set manually in our analysis. The spin excitations around this fixed ground state were then calculated using \textsc{SpinW} and the resulting spectra at all selected fields in our experimental range ($B = 0$-11~T) are shown as the orange lines in Fig.~3 of the main text.

For completeness, in \ref{dos2m} we show the densities of two-magnon states computed using the linear SWT one-magnon branches at fields of 0, 2, 3, 4, 5 and 8~T. The full two-magnon spectral functions are reweighted versions of these densities of states, which for the Y and V phases are more complex to compute and to normalize to the one-magnon branches. The density of states therefore serves as a useful semi-quantitative guide to the location of the two-magnon continuum in wavevector and energy, and to its qualitative shape. The densities of states in \ref{dos2m} each reflect the nine different continua arising from the three one-magnon branches, whose overlap results in the edge structures that appear throughout the Brillouin zone. The primary difference between linear SWT and our INS and MPS results is clearly the absence of well defined one-magnon branches extending over most of the zone at all fields outside the UUD phase. While the two-magnon continua of linear SWT occupy the entire zone over a broad energy window centred at 1 meV, they are in general too flat (except at 8 T), too uniform and have too many edges to bear a close resemblance to the continuum features in our measured and calculated spectral functions.

\section{Magnetization Response of the TLHAF}
\label{sec:mag}

To interpret the magnetic response despite the effects of the finite-temperature rounding, we estimate $B_{\mathrm{Sat}}$ from the TLHAF model parameters obtained by high-field INS (\ref{sec:lswt}A) and perform a grand canonical DMRG calculation of the full magnetization curves that allows us to deduce the boundaries of the 1/3-magnetization plateau.

\begin{figure}[t]
\centering
\includegraphics[width=0.9\linewidth]{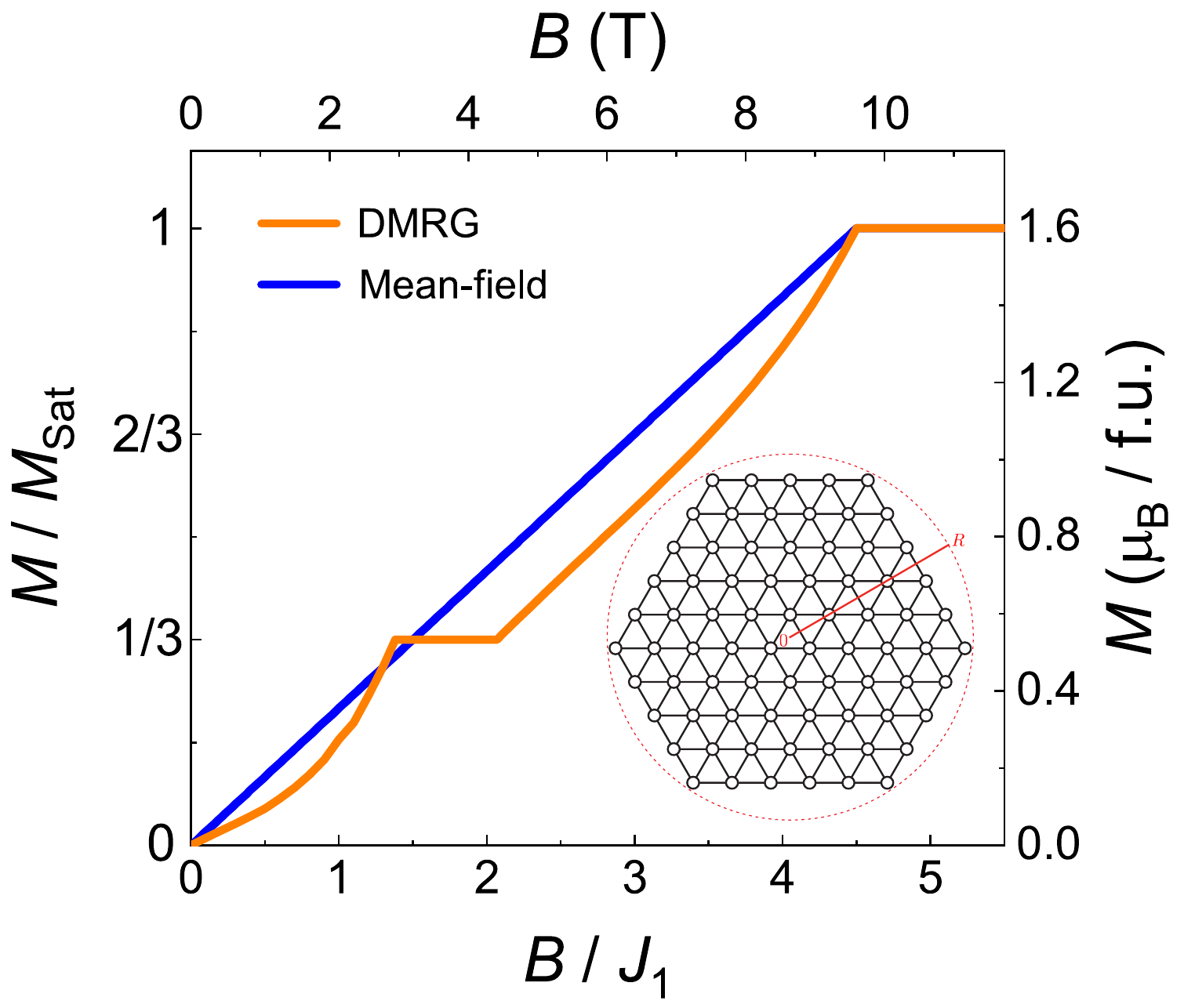}
\caption{Magnetization, $M(B)$, of the $J_1$-$J_2$ TLHAF with $J_2$/$J_1$ = 0.03 calculated by grand canonical DMRG and compared with the mean-field result. The inset shows the 75-site open cluster used for the DMRG calculation.}
\label{DMRG_MH}
\end{figure}

For a state-of-the-art numerical determination of the magnetization response of the $S = 1/2$ TLHAF, we apply the grand canonical DMRG method~\cite{Hotta2013}. This technique computes the infinitesimally small magnetization response to a change in the applied field, and thus the physical quantities it provides mimic the thermodynamic limit to approximately 1 part in $10^{3}$ for a 2D system. The method is based on a graded division of a finite-sized cluster into a centre region and edge regions, such that the centre reproduces the continuous bulk response by using the near-zero-energy states of the edge as a type of buffer. In a system of fixed size and shape, the grading is introduced by modulating the energy scale with the externally imposed function
\begin{align}
f(r) = \frac{1}{2} \left[ 1 + \cos\Big( \frac{\pi r}{R} \Big) \right],
\label{polarf}
\end{align}
which deforms the Hamiltonian smoothly from its standard energy at the centre of the system ($r = 0$) to zero at the open cluster edges ($r = R$). After obtaining the lowest eigen-wavefunction of the deformed Hamiltonian, the magnetization can be read as $M = \langle S^z (r = 0) \rangle$, because this wavefunction optimizes the expectation value to its thermodynamic limit at any given magnetic field. To mimic the results of a bulk measurement, appropriately weighted for the different possible directions of the magnetic field with respect to the triangular lattice, we performed our calculations using the hexagonal 75-site cluster shown in the inset of \ref{DMRG_MH}. The calculations were performed for the $J_1$-$J_2$ TLHAF using the interaction parameters deduced in \ref{sec:lswt}A and an in-plane $g$-factor of $g_{ab} = 3.2$, as deduced from ESR and INS in the main text. The result shown in \ref{DMRG_MH} is that reproduced in Fig.~1c of the main text, and was used for our extraction of the lower and upper boundaries ($B_{\rm l}$ and $B_{\rm u}$) of the 1/3 plateau.

\begin{figure}[t]
\includegraphics{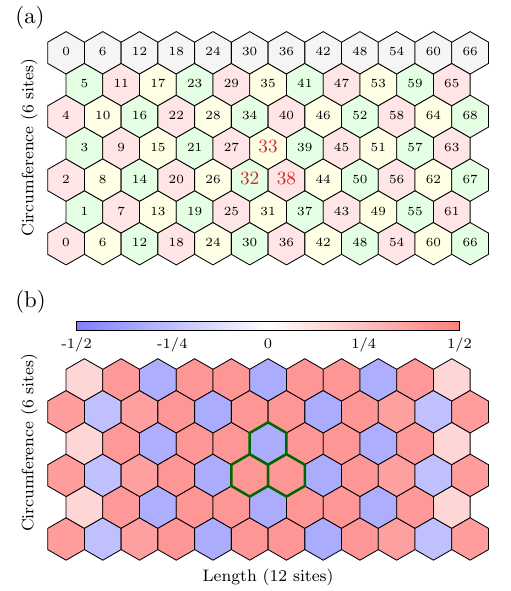}
\caption{(a) Example of the cylinder geometry used in our MPS calculations. The three sublattices are indicated by different colours. Numbers denote the index in the linear arrangement of the MPS and the three centre sites (used as initial excitation sites for the time evolution) are marked by red indices. Grey cells illustrate the wrapping specified by the periodic (XC) boundary conditions. (b) Local magnetization $\braket{S_i^z}$ measured in the ground state within the $1/3$ magnetization plateau, showing up-up-down order. The three centre sites are marked by green boundaries.}
\label{si_fig_mps1}
\end{figure}

\section{MPS Calculations}

The wavefunction of the spin system is represented by a matrix-product state (MPS), which assigns a tensor to each lattice site. The accuracy of the MPS approximation is controlled by the maximum bond dimension, $\chi$, of these tensors. To study the TLHAF we tested values of this parameter up to $\chi = 1024$, and will illustrate its role below. We used the Python package TenPy \cite{hauschild2018tenpy}, specifically its algorithms for the ground state and for time-evolution, as well as its support for preserving the U(1) symmetry of a model.

\begin{figure*}
\includegraphics{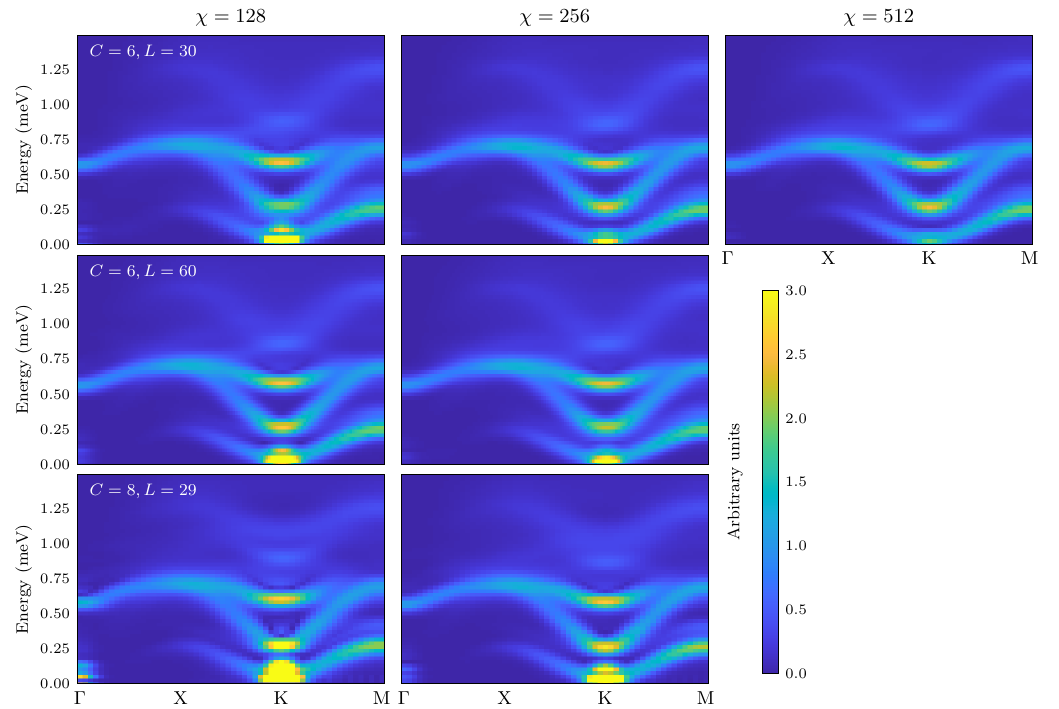}
\caption{Comparison of spectral functions computed using the MPS method for different system sizes and bond dimensions at the fixed values of the Gaussian envelope parameters stated in the text. Here we show the quantity $S_{xx}(\mathbf{Q},\omega) + S_{zz}(\mathbf{Q}, \omega)$ at a magnetic field corresponding to the experimental value $B = 3$~T. The columns show increasing values of the bond dimension, $\chi$, while the rows show cylinders of different sizes, as labelled in the panels in the first column.}
\label{fig-si-mps2}
\end{figure*}

We start by describing the cylinder geometry used for MPS calculations. We implement a cylinder with circumference $C$, length $L$ and the ``XC" boundary conditions shown in \ref{si_fig_mps1}(a)~\cite{szasz2020chiral,scheie2021}. This geometry provides a high momentum resolution along the Brillouin-zone path $\Gamma$-K-M that is of primary interest for the comparison with experiment. In our MPS calculations we determine the time-dependent spin-spin correlation function
\begin{align}
\begin{split}
\label{eq-si-mps1}
C&^{\alpha\beta}_\mathbf{r}(\mathbf{x}, t)
 = \braket{\hat{\mathcal{S}}^\alpha_{\mathbf{r} + \mathbf{x}}(t)\hat{\mathcal{S}}^\beta_\mathbf{r}(0)} \\
&= \braket{[\hat{S}^\alpha_{\mathbf{r}+\mathbf{x}}(t) - \braket{\hat{S}^\alpha_{\mathbf{r}+\mathbf{x}}(t)}]
[\hat{S}^\beta_\mathbf{r}(0) - \braket{\hat{S}^\beta_\mathbf{r}(0)}]} \\
&= \braket{\hat{S}^\alpha_{\mathbf{r}+\mathbf{x}}(t)\hat{S}^\beta_\mathbf{r}(0)} - \braket{\hat{S}^\alpha_{\mathbf{r}+\mathbf{x}}(t)} \braket{\hat{S}^\beta_\mathbf{r}(0)}
\end{split}
\end{align}
where the disconnected part is subtracted in order to remove the Bragg peaks in $C^{zz}$. Using $\ket{0}$ to denote the ground state and $E_0$ for the ground-state energy of $H$, we rewrite Eq.~\eqref{eq-si-mps1} in the form
\begin{align}
\label{eq-si-mps2}
C^{\alpha\beta}_\mathbf{r}(\mathbf{x}, t) &= \braket{0|\hat{S}^\alpha_{\mathbf{r}+\mathbf{x}} \ e^{-i(H-E_0)t} \ \hat{S}^\beta_\mathbf{r}|0} \nonumber\\
& \qquad - \braket{0|\hat{S}^\alpha_{\mathbf{r}+\mathbf{x}}|0}\braket{0|\hat{S}^\beta_\mathbf{r}|0},
\end{align}
where $\mathbf{r}$ is the site at which the initial spin operator is applied and $\mathbf{x}$ is the vector separation in the two-point correlator. Because of the residual U(1) symmetry of the Heisenberg Hamiltonian in a field, the total magnetization in the $z$ direction is conserved under time evolution. We take advantage of this symmetry to increase the performance of the calculation by restricting the MPS to a single magnetization sector. It follows that only the correlators which satisfy $\alpha\beta \in \{zz, +-, -+\}$ are non-zero. The other spin-spin correlation functions can be obtained from the identity $C^{xx}_\mathbf{r}(\mathbf{x}, t) \equiv C^{yy}_\mathbf{r}(\mathbf{x}, t) \equiv \frac{1}{2}[C^{+-}_\mathbf{r}(\mathbf{x}, t) + C^{-+}_\mathbf{r}(\mathbf{x}, t)]$.

\begin{figure*}[t]
\includegraphics[width=0.92\textwidth]{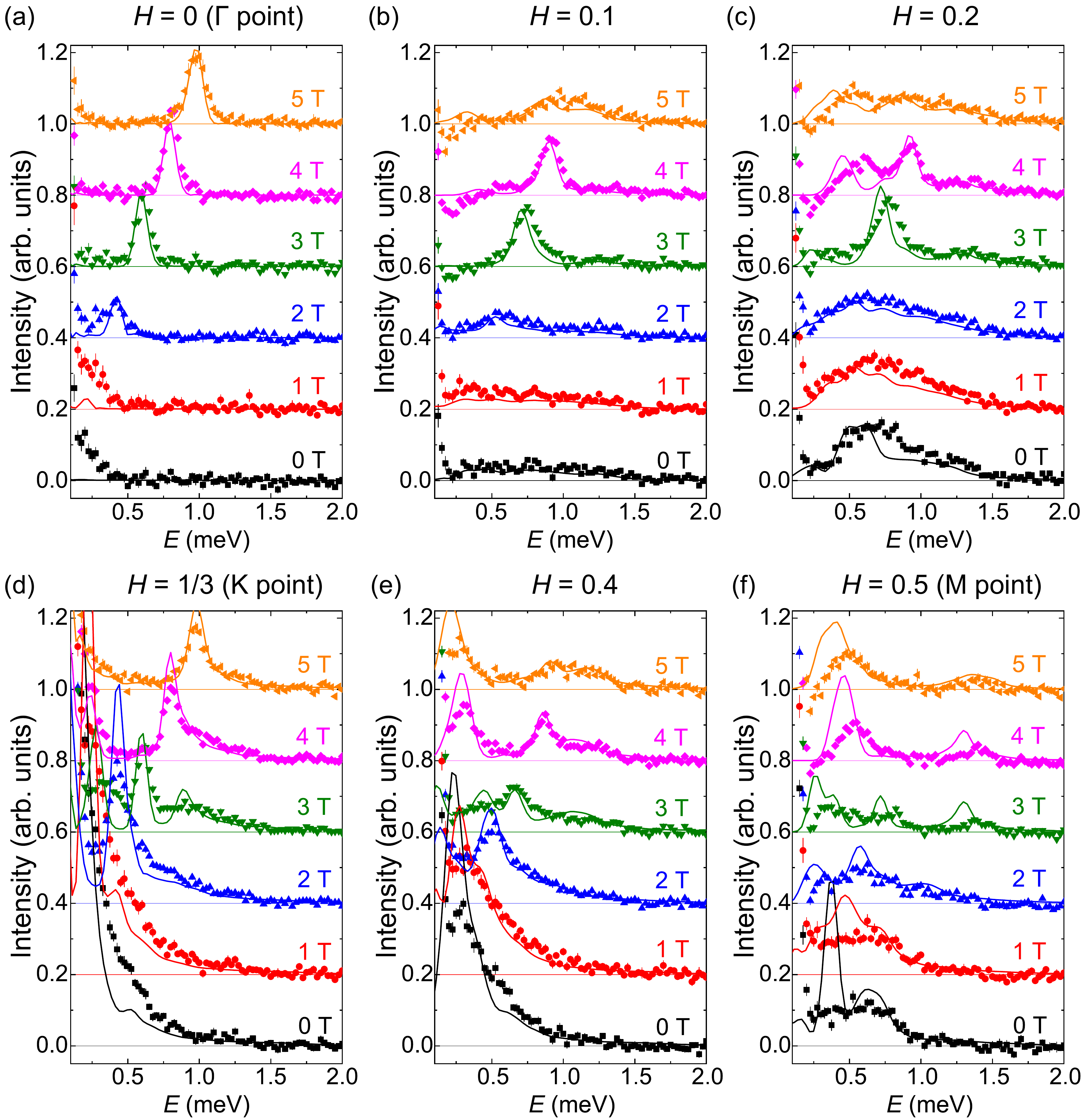}
\caption{Constant-$\mathbf{Q}$ cuts through the experimental (symbols) and calculated (solid lines) spectral functions for six different magnetic fields and six $\mathbf{Q}$ values along the [$H$~$H$~$2.5$] line. For clarity the cuts are shown with a vertical offset of 0.2.}
\label{Ecut}
\end{figure*}

In a typical time-evolution process, one selects a single site, $\mathbf{r}$, at the centre of the cylinder and uses a single time-evolved state to obtain the spectral function. However, the physics of the TLHAF in a field is such that there is a UUD-type symmetry-breaking of the $z$-axis magnetization at all finite fields. Our MPS set-up is therefore chosen such that the ground state in a field is a representative symmetry-broken state with a pronounced UUD-type local $z$-axis magnetization pattern, as shown in \ref{si_fig_mps1}(b). In fact it is more efficient to repeat the dynamical evolution using three distinct starting sites than it is to use a single MPS with all three symmetry-broken states superposed in a spatially uniform way. The reason for this is that the approximate tripling of the bond dimension required to represent this symmetric MPS would lead to a much more expensive time evolution, because of the expected $\chi^3$ scaling, than using three different starting states at a lower bond dimension does.

\begin{figure*}[t]
\center{\includegraphics[width=0.96\textwidth]{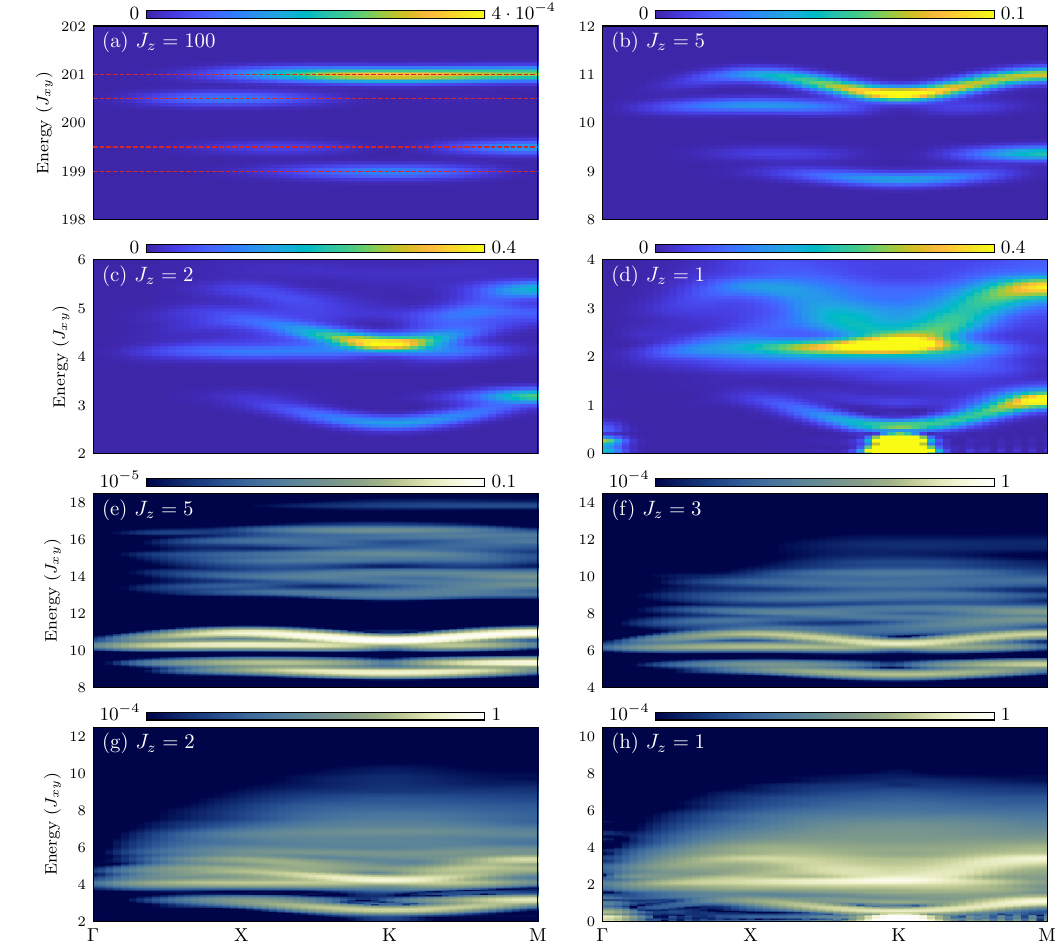}}
\caption{Cylinder MPS calculations of the longitudinal spectral function for the nearest-neighbour $S = 1/2$ TLAF with XXZ spin interactions. (a-d) Two-magnon bound states obtained with anisotropies $J_z/J_{xy} = 100$ (a), 5 (b), 2 (c) and 1 (d). We draw attention to the energy scales on the $y$-axes: the centre of the bound-state spectrum is set by $J_z$ and the width by $J_{xy}$. Dashed red lines in panel (a) show the four separate discrete levels obtained in the Ising limit. (e-h) Bound and scattering states shown on a logarithmic intensity scale for the anisotropy ratios $J_z/J_{xy} = 5$ (e), 3 (f), 2 (g) and 1 (h).}
\label{UUD_BoundState}
\end{figure*}

To obtain the spectral functions that correspond to the experiment, it is necessary to restore the lost translation symmetry. To this end, the spectral function $S_{\mathbf{r}, \alpha\beta}(\mathbf{Q}, \omega)$ in Eq.~(3) of the main text is calculated for the three sites $\mathbf{r}$ corresponding to the three sublattices of the central unit cell of a single symmetry-broken ground state (\ref{si_fig_mps1}) and we take the equally weighted average to obtain the symmetric spectral function
\begin{equation}
\label{eq-si-mps3}
S_{\alpha\beta}(\mathbf{Q}, \omega) = \frac{1}{3} \sum_{\mathbf{r}} S_{\mathbf{r},\alpha\beta}(\mathbf{Q}, \omega).
\end{equation}
This procedure is repeated for all $\alpha\beta \in \{zz, +-, -+\}$, meaning that in total nine time-evolution runs are required at every value of the magnetic field, $B$.

In addition, symmetries of the Heisenberg Hamiltonian allow us to obtain the negative-time correlation functions at no additional computational cost as the complex conjugate of the positive-time ones, i.e.~$C^{\alpha\beta}_\mathbf{r}(\mathbf{x}, -t) = \overline{C^{\alpha\beta}_\mathbf{r}(\mathbf{x}, t)}$. Putting these together, the symmetric spectral function of Eq.~\eqref{eq-si-mps3} simplifies to the form
\begin{align}
\begin{split}
S_{\alpha\beta}(\mathbf{Q}, \omega) \! = & \frac{2}{3} \sum_{\mathbf{r}} \int_{0}^{\infty} \mathrm{d}t \sum_\mathbf{x} e^{-i \, \mathbf{x}\cdot\mathbf{Q}} \label{eq-si-mps4} \\
& \! \left[ \cos(\omega t) \, {\rm Re} \, C^{\alpha\beta}_\mathbf{r}(\mathbf{x}, t) \! - \! \sin(\omega t) \, {\rm Im} \, C^{\alpha\beta}_\mathbf{r}(\mathbf{x}, t) \right] \! , \nonumber
\end{split}
\end{align}
where the outer sum runs over the three centre sites. To compensate for the finite cylinder length and time-step series, it is standard before Fourier transformation to convolve the correlation function with a Gaussian envelope,
\begin{equation}
C^{\alpha \beta}_\mathbf{r}(\mathbf{x}, t) \rightarrow e^{-\sigma_t t^2} e^{-\sigma_x (\mathbf{e}_L \cdot \mathbf{x})^2} \ C^{\alpha \beta}_\mathbf{r}(\mathbf{x}, t),
\end{equation}
which results in an effective broadening of the spectral function. For this we used $\sigma_t = 0.005 J_1^2$ and $\sigma_x = 0.02/a^2$, where $a$ is the unit-cell size of the TL and $\mathbf{e}_L$ the unit vector along the cylinder axis, leading to the effective resolution in energy and momentum given in the Methods section of the main text.

In \ref{fig-si-mps2} we show spectra obtained using this procedure in the UUD (1/3-plateau) phase, at a magnetic field equivalent to $B = 3$~T, to illustrate the degree to which our calculations have converged to the spectral function of the infinite system. We observe that increasing the cylinder circumference from $C = 6$ to 8, the length from $L = 30$ to 60 or the bond dimension of the MPS from $\chi = 256$ to 512 make only minor changes to the quality of the spectra at the values of $\sigma_t$ and $\sigma_x$ chosen to match the experimental resolution. We comment that higher-resolution experimental data would require higher $C$, $L$ or $\chi$ values to achieve convergence, and that our present MPS studies allow a factor-2 reduction in $\sigma_t$ and $\sigma_x$ (meaning a factor of $\sqrt{2}$ better resolution) before artifacts appear at isolated ${\bf Q}$ values. All of the results shown in Figs.~1, 3 and 4 of the main text were obtained using $C = 6$ and $L = 30$, with $\chi$ set to 1024 for $B < 2$~T and to 512 for all higher fields. All correlation functions were evaluated up to a final time $t_\textrm{max} = 90/J_1$, with a time step of $\Delta t = 0.1/J_1$.

\section{Comparison of Measured and Calculated Spectral Functions}

For a quantitative comparison of the MPS calculations with the experimental INS data, we present our results in two different ways. In Figs.~3b, 3d, 3f, 3h, 3j and 3l of the main text we show the spectral functions calculated for fields of 0, 2, 3, 4, 5 and 8~T with the symbols from Figs.~3a, 3c, 3e, 3g, 3i and 3k overlaid, which shows that the energies and wavevectors of all the characteristic spectral features are reproduced with quantitative accuracy. In \ref{Ecut} we show constant-$\mathbf{Q}$ cuts through the measured and calculated spectral functions for a range of $\mathbf{Q}$ points and applied magnetic fields. This extensive comparison reveals that the MPS calculations do an excellent job for all $\mathbf{Q}$ points at $B \ge 2$~T, losing their quantitative accuracy in locating the excitation features only on approaching the low-field limit ($B = 0$ and 1~T). Regarding the calculated intensities, we find that the MPS calculations achieve quantitative accuracy at low $|\mathbf{Q}|$ and between K and M, but clearly experience some challenges with the low-energy features near the K and M points.

\section{Bound States of Spin Waves in the UUD Phase}

With the goal of understanding the excitation spectrum in the UUD phase of the $S = 1/2$ TLAF, we focus on excitations at the same total $S^z$ as the 1/3-magnetization plateau. In order to elucidate the spectral response at the Heisenberg point, we consider an XXZ Hamiltonian with the magnetic field orientated parallel to the anisotropy axis,
\begin{equation}
\mathcal{H} = \sum_{\langle i,j \rangle} \! {\textstyle \frac12} J_{xy} \! \left( S^+_i S_j^- \! + \! S^-_i S_j^+ \! \right) + J_z S^z_i S^z_j - h \! \sum_i \! S^z_i,
\end{equation}
where $h = \mu_B g_{ab} B$. At $m = 1/3$, the ground state in the Ising limit, $J_z/J_{xy} \gg 1$, is the UUD product state illustrated in Fig.~4c of the main text. An excitation at the same $S^z$ consists at least of one pair of local spin-flips, $U \rightarrow D$ and $D \rightarrow U$. If the two flipped spins are spatially well separated, they cost an energy $3 J_z$ in the Ising limit, but if they are nearest neighbours then they cost only $2 J_z$. We note that, because the total $S^z$ is unchanged by two opposing spin flips, the $z$-axis magnetic field has no effect on the energies of these two-spin excitations.

As the next step we consider the perturbative limit with a finite, but small, $J_{xy}$. The nearest-neighbour bound states now show a spatially localized yet partially mobile structure. A schematic representation is shown in Fig.~4c of the main text, where the flipped red spin is confined, but can hop around the flipped blue spin along a finite chain of six sites, i.e.~on the hexagon around the blue spin. The discrete energy spectrum of this six-site chain shows a lifting of the bound-state degeneracy to six flat bands with energies $\{2J_z + J_{xy}, 2J_z + J_{xy}/2, 2J_z - J_{xy}/2, 2J_z - J_{xy}\}$, with the second and third energies each twofold degenerate. The manifold of excitations arising from pairs of flipped spins far away from each other is also split at first order in $J_{xy}$, and forms a two-particle continuum at an energy around $3J$ whose support can be derived from the dispersion of the single spin-flips.

The next step is to determine whether this structure of bound states is visible in the longitudinal dynamical structure factor, $S_{zz}(\mathbf{Q},\omega)$. For this we performed MPS calculations for the nearest-neighbour model at different values of $J_z/J_{xy}$, and in \ref{UUD_BoundState}(a) we show the spectral function at $J_z/J_{xy} = 100$ along the usual $\Gamma$-K-M path in the Brillouin zone. Indeed one observes that the spectral weight becomes $\mathbf{Q}$-dependent, and that the excitation energies are confined to the perturbative branches (horizontal dashed red lines) discussed above. As $J_z/J_{xy}$ is lowered to $5$ [\ref{UUD_BoundState}(b)], $2$ [\ref{UUD_BoundState}(c)] and the Heisenberg point [\ref{UUD_BoundState}(d)], the bound-state branches acquire an increasing dispersion and broaden in energy; in particular, while the lower bound state remains relatively sharp as $J_z/J_{xy} \rightarrow 1$, the upper scattering states broaden into the bow-tie structure of continuum IV.

To track the origin of this broadening, in \ref{UUD_BoundState}(e-h) we show four spectral functions on a logarithmic intensity scale. In \ref{UUD_BoundState}(e), which corresponds to \ref{UUD_BoundState}(b), the two-magnon scattering continuum begins to become visible, centred at an energy around $3J_z$ ($15J_{xy}$). As $J_z/J_{xy}$ is lowered, this continuum rises in intensity and overlaps increasingly with the energy window of the bound states, until at the Heisenberg point [\ref{UUD_BoundState}(h)] only a part of the lowest bound-state branch remains as the isolated and well defined mode observed in experiment (Fig.~4a of the main text).